\documentclass[fleqn,12pt]{wlscirep}
\usepackage[utf8]{inputenc}
\usepackage[T1]{fontenc}
\usepackage{rotating, graphicx}
\usepackage{wrapfig}

\usepackage{lineno}
\usepackage{setspace}

\usepackage[finalnew]{trackchanges}
\addeditor{Rev1}

\usepackage{soul}
\soulregister\ref7
\soulregister\cite7
\soulregister\citeA7
\soulregister\citeyear7
\soulregister\citeauthor7
\soulregister\url7

\title{Deep Neural Networks for Active Wave Breaking Classification}

\author[1,*]{Caio E. Stringari}
\author[1,2,+]{Pedro V. Guimar\~aes}
\author[1,+]{Jean-Fran\c{c}ois Filipot}
\author[1,]{Fabien Leckler}
\author[1,]{Rui Duarte}
\affil[1]{France Energies Marines, Plouzané, 29280, France}
\affil[2]{PPGOceano, Federal University of Santa Catarina, Florianópolis, 88040-900, Brazil}

\affil[*]{Caio.Stringari@france-energies-marines.com}

\affil[+]{these authors contributed equally to this work}

\begin{abstract}
Wave breaking is an important process for energy dissipation in the open ocean and coastal seas. It drives beach morphodynamics, controls air-sea interactions, determines when ship and offshore structure operations can occur safely, and influences on the retrieval of ocean properties from satellites. Still, wave breaking lacks a proper physical understanding mainly due to scarce observational field data. Consequently, new methods and data are required to improve our current understanding of this process. In this paper we present a novel machine learning method to detect active wave breaking, that is, waves that are actively generating visible bubble entrainment in video imagery data. The present method is based on classical machine learning and deep learning techniques and is made freely available to the community alongside this publication. The results indicate that our best performing model had a balanced classification accuracy score of $\approx$ 90\% when classifying active wave breaking in the test dataset. An example of a direct application of the method includes a statistical description of geometrical and kinematic properties of breaking waves. We expect that the present method and the associated dataset will be crucial for future research related to wave breaking in several areas of research, which include but are not limited to: improving operational forecast models, developing risk assessment and coastal management tools, and refining the retrieval of remotely sensed ocean properties. 

\end{abstract}

\begin{document}

\flushbottom
\maketitle

\thispagestyle{empty}

\doublespacing

\section*{Introduction}

Wave breaking is one of the most challenging water wave phenomena to investigate. Despite nearly two centuries of research, the current practical description of the process is still mostly empirical \cite{Battjes1978, Thornton1983, Banner2000, Banner2002}.\remove[Rev1]{Since Miche \cite{Miche1944b} it has been well established that regular waves over a flat bottom will start to break if they overcome a certain steepness threshold, that is, the ratio between wave height ($H$) and wavelength ($L$). This criterion does not, however, hold for irregular waves over complex bathymetry. Such waves break when the horizontal velocity in the crest reaches, or is close to, the wave phase speed \cite{Barthelemy2018, Derakhti2020}.}
\add[Rev1]{Precise} wave breaking \add[Rev1]{modelling} requires\remove[Rev1]{, therefore, to} explicit\change[Rev1]{ly}{ knowledge} \change[Rev1]{resolve}{of} the wave phase \add[Rev1]{speed} and the fluid velocity distribution on the \add[Rev1]{wave} crest \add[Rev1]{which can only be done }by numerically solving the Navier-Stokes equations in a framework that\remove[Rev1]{properly accounts for small turbulence fluctuations, thus precisely describing the process of wave breaking \cite{Lubin2015, Seiffert2017}.} \remove[Rev1]{Such approaches are extremely}\add[Rev1]{currently is too} computationally expensive \add[Rev1]{for practical applications}\cite{Cavaleri2006}. \change[Rev1]{ and, therefore,}{Consequently,} current large-scale state-of-the-art wave models rely on statistical approaches (mostly the spectral approach) to represent the waves \cite{WW32019, Booij1999, THEWANDIGROUP1988}. This type of models parameterizes wave breaking as a function of known parameters such as the wind speed \cite{WW32019}, the local wave height to water depth ratio \cite{Booij1999} or semi-empirical breaking wave height probability distributions \cite{Thornton1983, Filipot2010a}. \add[Rev1]{Due to a lack of observed data, the constants involved in these models have been derived from limited datasets that may not adequately represent the natural environment. For example, a recent study has shown that popular surf zone parametric wave breaking models incorrectly represented the fraction of broken waves in their formulations with errors >50\%, despite these models being able to adequately represent surf zone energy dissipation possibly due to parameter tuning \cite{Stringari2019a}. This paper aims to start addressing the data unavailability issue by providing a reliable and reproducible method to detect and track waves that are actively breaking in video imagery data.}

\change[Rev1]{Wave breaking is also directly correlated directly correlates to ocean-atmosphere interactions\cite{Reul2003, Kleiss2011, Romero2012, Zappa2012, Gemmrich2013}. Models that consider}{Wave breaking directly affects several environmental phenomena. For example,}  wave breaking \add[Rev1]{is considered }as the main driver for air-sea exchanges\cite{Melville2002},\remove[Rev1]{mostly exclusively} \add[Rev1]{being often }describe\add[Rev1]{d} \change[Rev1]{air-sea fluxes based}{as a function} of Phillips' \cite{Phillips1985} $\Lambda(c)dc$ parameter. This parameter is defined as the \change[Rev1]{expected average density length of breaking crests moving with velocity in the range $c$ to $c+dc$ per unit of surface area}{average total length per unit surface area of breaking fronts that have velocities in the range $c$ to $c+dc$} and its moments are assumed to correspond to different phenomena such as whitecap coverage (second moment), rate of air entrainment (third moment), and momentum flux (fourth moment). \change[Rev1]{However, a}{A}s previously identified \cite{Banner2014}, different interpretations of how data is processed to obtain $\Lambda(c)dc$ resulted in differences of 175\% in the first moment and 300\% in the fifth moment of $\Lambda(c)$ \cite{Gemmrich2008, Banner2014}. \change[Rev1]{Nevertheless, t}{T}his issue does not \add[Rev1]{seem to} limit the applicability of Phillips' framework. A recent study, for example, used a wave breaking parameterization fully based on $\Lambda(c)dc$ to derived whitecap coverage off the coast of California\cite{Romero2019}.\change[Rev1]{If more robust wave breaking detection methods and larger datasets confirm the validity of Phillips' framework (or are used to improve its current form), approaches such as Romero's\cite{Romero2019} could lead to much improved coupled ocean-atmosphere models.}{ To the authors' knowledge, however, no study has assessed the impact that errors in wave breaking detection has on measured $\Lambda(c)$ distributions and how such errors would impact on the conclusions that have been drawn  from these models (for example, the assumption that $\Lambda(c)$ is a simple function of wind speed\cite{Melville2002}). More importantly, in the case which new observations diverge from Phillips' original theory (as we will later show in this paper), new and improved wave breaking models will certainly follow.}

Further, breaking waves have a direct impact on the remote sensing of the ocean surface from satellites\change[Rev1]{ or other}{,} aircraft\add[Rev1]{, or platform mounted instruments}. Wave breaking has been \change[Rev1]{shown}{observed} to significantly modulate measurements of backscatter off radar\change[Rev1]{ and satellites}{s} with large increases in backscatter \change[Rev1]{being observed in direct association}{being directly correlated} with breaking waves\change[Rev1]{, both in field and laboratory experiments \cite{Jessup1990, Trizna1996, Ericson1999}}{\cite{Yurovsky2018}. Given that part of the data that will be used here (see Methods for details) coincides with the location of previous wave backscatter studies\cite{Yurovsky2018}, data derived from our method could to lead to explicit formulations correlating wave breaking and radar backscatter. Such formulations could, in the future, lead to more reliable satellite-derived scatterometer data, which currently ignores \cite{Huang2017} or relies on empirical wave breaking models\cite{Hwang2010}}. The foam generated by breaking waves at the sea surface also \change[Rev1]{significantly contributes to}{influences on the measurement of} microwave brightness temperature signatures for moderate to high wind speeds \cite{Monahan2002, Reul2003}. To date, no study has, however, systematically quantified such temperature signatures for large field datasets and most of our knowledge \change[Rev1]{of this phenomena is purely theoretical or} is empirical. \remove[Rev1]{New precise measurements of wave breaking will help on validating the current backscatter models or could be used to propose new formulations.}\add[Rev1]{Such limitation could be addressed by extending wave breaking detection and tracking methods to infrared wavelengths, which is directly correlated to water temperature\cite{Carini2015}. Advancements in this regard could lead to improvements in global climate models given that the inclusion of wave-generated heat transfer has shown to reduce cold biases in such models for non-breaking waves\cite{Wang2019}. Breaking waves are expected to transfer more heat than non-breaking waves\cite{Komori2018} and, when considered, could improve climate models even further}.

\remove[Rev1]{A major limitation underlying all the research described above has been the small number of detailed field observations of breaking waves and their associated statistics, for example, wave breaking length, duration and speed. As a consequence of this limitation, the constants involved in all models that depend on wave breaking have been derived from a very limited set of sea state conditions which may not adequately represent the natural environment. For example, a recent study has shown that popular surf zone parametric wave breaking models incorrectly represented the fraction of broken waves in their formulations, despite these models being able to adequately represent surf zone energy dissipation \cite{Stringari2019a}. Further, as briefly mentioned before, different methods have been proposed to detect wave breaking and to compute the associated wave statistics. In deep water, at least three \cite{Banner2014} different approaches have been used to obtain Phillips' $\Lambda(c)dc$ \cite{Phillips1985} distribution, which resulted in the measurements either agreeing or disagreeing with Phillips' original theory, as mentioned above. Finally, despite being the same physical process, wave breaking in deep and shallow water have been treated as two independent phenomena in both modelling and field research fronts \cite{Kleiss2010, Melville2002, Sutherland2013, Schwendeman2014, Stringari2019, Almar2014, Yoo2010} with very few modelling exceptions \cite{Filipot2010a, Filipot2012}.}

With the recent explosion in the usage of machine learning, particularly deep neural networks, it was only a matter of time before these techniques were adapted to wave research. Recent studies have shown that deep neural networks can accurately be used to classify different types of surf zone breakers \cite{Buscombe2019}, obtain wave heights from video data \cite{Buscombe2020}, track waves in the surf zone\cite{Kim2020} and the shoreline position \cite{Bieman2020} and, when applied to pressure transducer data, distinguish between broken and unbroken waves \cite{Stringari2019a}. In this paper, we describe a robust and extensible method to identify and track breaking waves in video imagery data using a combination of classic machine learning algorithms and deep learning. More importantly, we make the present dataset and code base fully available for future researchers who can either use it directly, re-train the models adding more training data, or expand the present method to other cases (for example, object segmentation). The development and standardization of a general framework to detect wave breaking in video imagery data should help to provide the wave breaking statistics database that is currently needed to assess, or further develop, our current understanding of several processes related to breaking waves. Ultimately, the efforts initiated here aim to lead to improvements in several areas of research such as wave modelling and forecasting, \add[Rev1]{wave energy tracking and harvesting \cite{Zheng2019}}, ocean-atmosphere interaction \cite{Romero2019}, remote sensing of the ocean\cite{Yurovsky2018}, and safety at sea and coasts\cite{Filipot2019}. \remove[Rev1]{The major advantage of machine learning models \cite{Buscombe2020, Bieman2020} over process-based approaches \cite{Stringari2019, Kleiss2010, Sutherland2013, Mironov2008} is that machine learning models are able to generalize on new data much better than process-based methods. For example, pixel intensity thresholding \cite{Mironov2008, Schwendeman2014} and color\cite{Stringari2019, Hoonhout2015} based  algorithms for wave breaking detection must be laboriously adapted to each new dataset. Furthermore, process-based approaches make it difficult to combine data from different sources into larger datasets because the algorithms used to process these data are hand-crafted for specific uses. Fewer process-based approaches\cite{Mironov2008} tried to remove user dependence but ended up failing to generalize for all conditions. Machine learning approaches, on the contrary, have the advantage of being able to adapt to new data even for small training datasets given their very nature\cite{Olson2018}.}

\remove[Rev1]{In this paper, we describe a robust and extensible method to identify and track breaking waves in video imagery data using a combination of classic machine learning algorithms and deep learning. More importantly, we make the present dataset and code base fully available for future researchers who can either use directly, re-train the models adding more training data, or expand the present method to other cases (for example, object segmentation). To the authors' knowledge no fully open-source method of this kind currently exists. The development and standardization of a general framework to detect wave breaking in video imagery data should help to provide the wave breaking statistics database that is currently needed to assess, or further develop, our current understanding of several process related to breaking waves. Ultimately, the efforts initiated here aim to lead to improvements in several areas of research such as wave modelling and forecasting, ocean-atmosphere interaction, remote sensing of the ocean, and safety at sea and coasts.}

\section*{Method}

\subsection*{Model Definition}

In this study, we have developed a novel method to detect active wave breaking in video imagery data. Similarly to previous studies, we exploit the fact that breaking wave crests generate characteristic white foam patches from bubble entrainment \cite{Mironov2008,Sutherland2013}. Differently from \add[Rev1]{the vast majority of} previous methods, here we make a clear distinction between active wave breaking (that is, visible foam being actively generated by wave breaking) from passive foam and to all other non-wave breaking instances. \add[Rev1]{To the authors' knowledge, only the methods of Mironov \& Dulov\cite{Mironov2008} and Kleiss \& Melville\cite{Kleiss2010} have previously attempted to distinguish between active wave breaking and passive foam. Both their methods start similarly to ours (see next paragraph) by identifying bright pixels in a given series of images via thresholding. Next, interconnected pixels are connected in space and time using a nearest neighbor search (Mironov \& Dulov\cite{Mironov2008}) or by extracting connected contours (Kleiss \& Melville\cite{Kleiss2010}). In both previous methods, the separation between active and passive wave breaking is then done by empirically defining foam expansion direction and velocity limits. The present method aims to remove empirical steps in active wave breaking detection by using data-driven methods and deep neural networks instead of experiment-specific parameters.} \remove[Rev1]{Furthermore, the present method provides a measurement of the active wave breaking classification error which, to the authors' knowledge, has never been reported (see below for further details).}

\change[Rev1]{Firstly,}{Our method starts by na\"ively identifying} wave breaking candidates \remove[Rev1]{were na\"ively identified}in a given input image. This was done by identifying bright (that is, white) pixels in the image using \change[Rev1]{thresholding algorithms}{the local thresholding algorithm} available from the \textit{OpenCV}\cite{OpenCV} library \add[Rev1]{with its default parameters}. Interconnected regions of bright pixels in the image were then clustered using DBSCAN \cite{Ester1996} and a minimum enclosing ellipse \cite{Moshtagh2005} was fitted to each identified cluster. \add[Rev1]{The DBSCAN step requires the user to define a minimum number of pixels to be clustered (default value is 10 for all data used here) and a maximum distance allowed between pixels (default value of 5 for all data used here).} This step resulted in large quantities of bright pixel patches being detected. At this stage, it was not possible to determine whether a given identified cluster of pixels represented active wave breaking, passive foam, or any other instance of bright pixels (for example, birds, boats, coastal structures, sun glints or white clouds). Generically, this step can be thought of a feature extraction step and can be replaced by any other equivalent approach (for example, image feature extractors). To avoid  exponential memory consumption growth generated by DBSCAN, the input image may be subdivided into blocks of regular size. This step was easily parallelized with performance increasing linearly with number of computing threads.

\begin{figure}[htp]
    \centering
    \includegraphics[width=\linewidth]{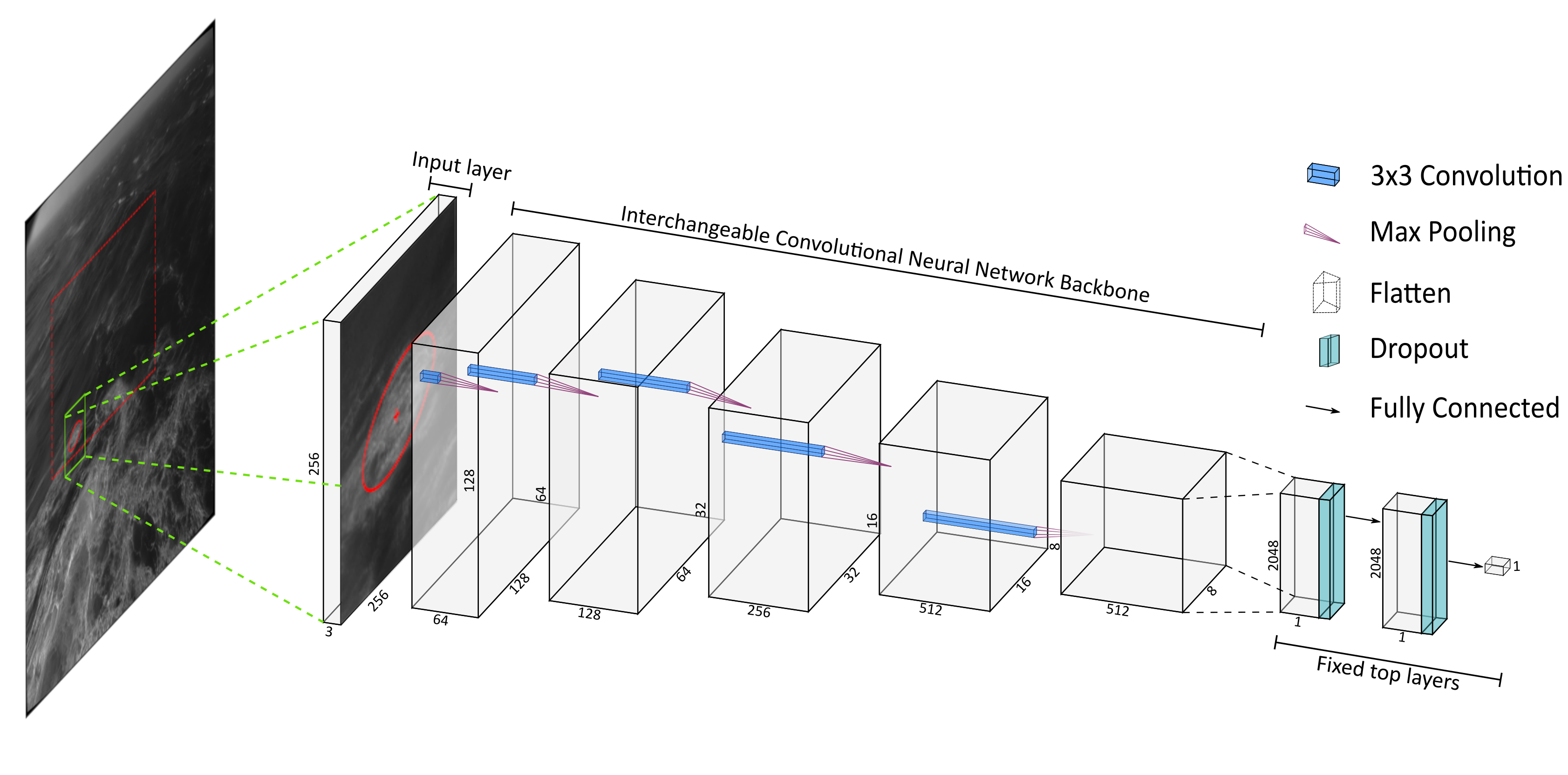}
    \caption{Schematic representation of a deep convolutional neural network. The input layer consists of an active wave breaking candidate and has shape 256x256x3 (image height, image width and number of RGB channels). In the case of a grey-scale image, the single grey channel was triplicated. The red dashed rectangle in the analyzed image shows the region of interest which, in this particular case, was equal to the stereo video reconstruction area. The intermediary convolutional layers (or backbones) are interchangeable with different options available (see text for details). The function of the convolutional layers is to extract features from the input image by using convolutions (\remove[Re1]{usually}3x3 \add[Rev1]{in this example}) and max pooling\cite{Nagi2011} (that is, selecting the brightest pixel in a given window). The last convolutional layer is flattened (that is, turned into a one-dimensional vector) and is connected to two fully-connected (that is, multi-layer perceptron-like) layers and one final classification layer. A 50\% dropout (that is, random selection of neurons in a layer) is applied after each fully connected layer. The final classification layer has one unit and uses sigmoid activation with a threshold of 0.5.}
    % \caption{Legend (350 words max). Example legend text.}
    \label{fig:architecture}
\end{figure}

The second step of the method consisted of training a deep convolutional neural network to distinguish between active wave breaking and passive foam (and all other non-desired occurrences). This step can be \change[Rev1]{thought of}{understood as} a binary classification step. Figure \ref{fig:architecture} shows a schematic representation of the present method. From the original images, randomly selected subsets of size 256x256 pixels centered on the fitted ellipses were extracted and manually labelled as either active wave breaking (label 1, or positive) or otherwise (label 0, or negative). The present training dataset was generated using raw video imagery data from Guimara\~es et al. \cite{Guimaraes2020} and consisted of 19000 training and 1300 testing images (see Table \ref{tab:dataset}). The training dataset is further split into 80\% training and 20\% validation data\cite{Hastie2009}. The validation dataset \change[Rev1]{was used}{is reserved} for \add[Rev1]{future} hyper-parameter fine-tuning. Note that the present dataset has a class imbalance of $\approx$90\% towards the negative label, that is, for each active wave breaking sample (label 1, or positive) there are nine instances that were not active wave breaking (label 0, or negative). Data augmentation \cite{Shorten2019} (rotation, vertical and horizontal flips, and zoom) was employed during training to increase the variety  of samples of the positive label.

Five state-of-the-art neural network architectures, or backbones (VGG16 \cite{Simonyan2015}, ResNet50V2 \cite{He2016a}, InceptionResNetV2 \cite{Szegedy2017}, MobileNetV2 \cite{Howard2017} and EfficientNetB5 \cite{Tan2019}), were implemented and can be chosen by end users (see \add[Rev1]{our Github code repository} \url{https://github.com/caiostringari/deepwaves} for usage guidance). \add[Rev1]{All of these backbones make use of convolutional layers to gradually extract information from the input images. The main differences between them are how many layers they have, the size of the convolution windows, how data is normalized in each layer and how each layer connects to each other (or to previous layers in the stack). For example, VGG16 is 16 layers deep, uses 3x3 convolution windows and has no normalization. InceptionResNetV2 is 164 layers deep, uses a mix of 5x5, 3x3 and 1x1 convolution windows and uses batch normalization\cite{Ioffe2015} to help avoiding overfitting. Residual Networks (namely, ResNet50V2) not only connect adjacent layers but also take into account errors (residuals) from previous layers. In general, more modern architectures (namely, EfficientNet) are wider (mainly by having parallel convolution windows) as well as much deeper than older architectures (namely, VGG16).} The final top layers of the network were fixed \add[Rev1]{for all backbones} and consisted of flattening the last convolutional layer, two fully-connected layers with 50\% dropout \cite{Srivastava2014}, and a final classification layer with sigmoid activation.

The optimization step (training) was done using the \textit{Adam} \cite{Kingma2014} implementation of the stochastic gradient descent method \cite{Ruder2016} and binary cross-entropy was used as the loss function. Note that this step must be computed using a graphics processing unit (GPU) in order to achieve feasible computation times. \add[Rev1]{The models took from three (VGG16) to twelve (EffientNetB5) hours to train using a NVIDIA GTX 1080 GPU with a batch size of sixty-four images}. After the neural networks were sufficiently trained, the best performing network (VGG16, see below) was used to classify all the na\"ively identified wave breaking candidates and only the events classified as active wave breaking were kept for further analyses. \add[Rev1]{Although VGG16 was chosen for presenting the results, the performance of the other architectures is nearly identical to VGG16 on real-world applications. Finally, note that user-tunable parameters for the neural networks (for example, learning rate and neuron activation thresholds) were kept unchanged from their default values in the TensorFlow library\cite{tensorflow2015}. This implicates that more aggressive parameter optimization could improve the results presented here even further. For sake of brevity, we refer the reader to the project's code repository for guidance on how to select hyper-parameters.}

\begin{table}[ht]
  \centering
  \caption{Data characterization summary table. $f$ is the sampling frequency in Hertz, $D$ is the duration of the experiment in seconds, $H_s$ is the significant wave height computed from the wave spectrum, $T_{p1}$ and $T_{p2}$ are, respectively, the peak wave period of the first and second spectral partitions computed following Hanson \& Jensen\cite{Hanson2004}, $D_p$ is the peak wave direction, and $U_{10}$ is the wind speed measured or converted (denoted by the $*$) to a height of 10m above the sea surface using Large \& Pond's\cite{Large1981} formula. Tr. S. and Ts. S. are the sample size of the train and test datasets, respectively.}
  \resizebox{\columnwidth}{!}{%
    \begin{tabular}{lllllllllll}
    \hline
    Location     & Date and Time      & $f$ $[Hz]$ & D. $[min]$ & $H_s$ {[}m{]} & $T_{p1}$ $[s]$ & $T_{p2}$ $[s]$ & $U_{10}$ $[ms^{-1}]$ & $D_p$ & Tr. S. & Ts. S. \\
    \hline
    Black Sea    & 2011/10/01 14:18   & 12             & 07             & 0.3        & 6.20           & 3.10          & 10.7*         & WSW      & 1000       & 100       \\
    Black Sea    & 2011/10/04 09:38   & 12             & 20             & 0.36       & 6.10           & 2.63          & 10.1*         & WSW      & 1000       & 100       \\
    Black Sea    & 2011/10/04 11:07   & 12             & 30             & 0.45       & 6.10           & 3.16          & 12.2*         & WSW      & 1000       & 100       \\
    Black Sea    & 2011/10/04 13:30   & 12             & 30             & 0.55       & 6.60           & 3.71          & 12.9*         & WSW      & 1000       & 100       \\
    Black Sea    & 2013/09/22 13:00   & 10             & 15             & 0.66       & 4.30           & 30            & 8.7*          & E        & 1000       & 100       \\
    Black Sea    & 2013/09/25 12:15   & 12             & 15             & 0.41       & 4.10           & 1.20          & 6.1*          & N        & 1000       & 100       \\
    Black Sea    & 2013/09/30 10:20   & 12             & 15             & 0.65       & 5.70           & 1.80          & 15.2*         & N        & 1000       & 100       \\
    Adriatic Sea & 2014/03/27 09:10   & 12             & 60             & 1.36       & 5.02           & -             & 9.9           & ENE      & 2000       & 100       \\
    Adriatic Sea & 2015/03/05 10:35   & 12             & 60             & 1.33       & 6.10           & 5.02          & 9.0           & ENE      & 2000       & 100       \\
    Yellow Sea   & 2017/05/13 05:00   & 10             & 05             & 1.93       & 5.20           & 4.02          & 13.4          & NW       & 2000       & 100       \\
    La Jument    & 2017/12/15 14:20 & 10             & 30             & 5.88       & 10.00          & 6.80          & 13.3          & W        & 2000       & 100       \\
    La Jument    & 2018/01/03 09:39   & 10             & 30             & 10.03      & 12.80          & 9.30          & 17.9          & W        & 2000       & 100       \\
    La Jument    & 2018/01/04 11:43   & 10             & 30             & 7.52       & 11.10          & -             & 14.7          & W        & 2000       & 100       \\
                 &                    &                &                &            &                &               &               &         \textbf{Total:} & 19000      & 1300     \\
    \hline
    \end{tabular}
    }%
  \label{tab:dataset}%
\end{table}%

The last step of the method consisted of grouping the active wave breaking events in space and time (note that at this stage bright pixels are only grouped in space). Time-discrete wave breaking events had their bounding ellipse region filled in pixel space ($i, j$) and were stacked in time ($t$) resulting in a point-cloud-like three-dimensional ($i,j,t$) structure. The DBSCAN\cite{Ester1996} algorithm was then used to cluster these data and obtain uniquely labelled clusters. The two parameters for the DBSCAN algorithm, that is, the minimum number of samples in each cluster ($n_{min}$) and the minimum distance allowed between two samples ($eps$), were set to be equals to the minimum number of points inside a fitted ellipse among all ellipses ($n_{min}$) and equals to the sampling frequency in Hertz ($eps$). These values for $n_{min}$ and $eps$ were constant \change[Rev1]{within}{among} the analyzed datasets. Note that this final step can be replaced by any other density-based clustering algorithm or other more sophisticated algorithms \add[Rev1]{such as SORT\cite{Bewley2016} (which is also available on the project's code repository but not was used here)}.

Note that up to the clustering step, all calculations were done in pixel domain and "time" was obtained from sequential frame numbers. To convert between pixel and metric coordinates, the grids constructed using the stereo-video dataset available from  Guimar\~aes et al. \cite{Guimaraes2020} were used in the present study. If stereo video data are not available, such conversions could be done knowing the camera position in real-world (that is, metric) coordinates, which is usually done using surveyed ground control points\cite{Holman2007,Schwendeman2014}. Conversion from sequential frame numbers to time (in seconds) can be done by knowing the sample rate (in frames per second, that is, Hertz) for the data. \add[Rev1]{The total amount of computational time required to process 20 minutes of raw video recorded at 10Hz with an image size of 5 megapixels is approximately two hours on a modern six-core computer (assuming that a pre-trained neural network is available). Much shorter processing times are achievable for smaller images sizes and higher number of computation threads.} 

\subsection*{Evaluation Metrics}

Due to the imbalanced characteristic of the dataset, the classification accuracy score in isolation is not an appropriated metric to evaluated the present test dataset. For instance, a classifier that guesses that all labels are negative (that is, 0) would automatically obtain a high score ($\approx$ 90\%). To properly assess the performance of the classifier, more robust metrics were defined. These metrics are defined as follows:

\begin{itemize}

    \item True Positives ($TF$) and True Negatives ($TN$)  are samples that were correctly classified. False Positives ($FP$) or Type I errors are false samples that were incorrectly classified as true, and False Negatives ($FN$) or Type II errors are true samples that were classified as false.
    
    \item Accuracy is the percentage of examples correctly classified considering both classes, that is,  Accuracy=$\frac{TP+TN}{T+N}$, where $T$ and $N$ are the total number of positive and negative samples, respectively.
    
    \item Precision is the percentage of predicted positives that were correctly classified, that is,  Precision=$\frac{TP}{TP+FP}$.
    
    \item Recall is the percentage of actual positives that were correctly classified, that is, Recall=$\frac{TP}{TP+FN}$ 
    
    \item Area under the curve ($AUC$) is the area defined by plotting the FP rate against the TP rate (also referred to as receiver operating characteristic curve). This metric indicates the probability of a classifier ranking a random positive sample higher than a random negative sample.

\end{itemize}

\noindent All metrics described above were monitored during the training process and were individually assessed to rank model performance. The training curves shown in Figure \ref{fig:training_and_cm}-a were used to assess when overfitting, that is, decreases in the loss function value for the training dataset that were not reflected in the validation dataset, started to occur. Training epochs after overfitting started were discarded.  Here we favor the $AUC$ curves as shown in Figure \ref{fig:training_and_cm}-b  to indicate better performing models because AUC is a more robust metric than the classification score and at the same time presents a smooth evolution with training epochs. Finally, a confusion matrix (or table of confusion) as shown in Figure \ref{fig:training_and_cm}-c was plotted for each model to assess Type I and Type II errors.

\section*{Results}\label{results}

\subsection*{Classifier performance}

\begin{figure}[htp]
    \centering
    \includegraphics[width=\linewidth]{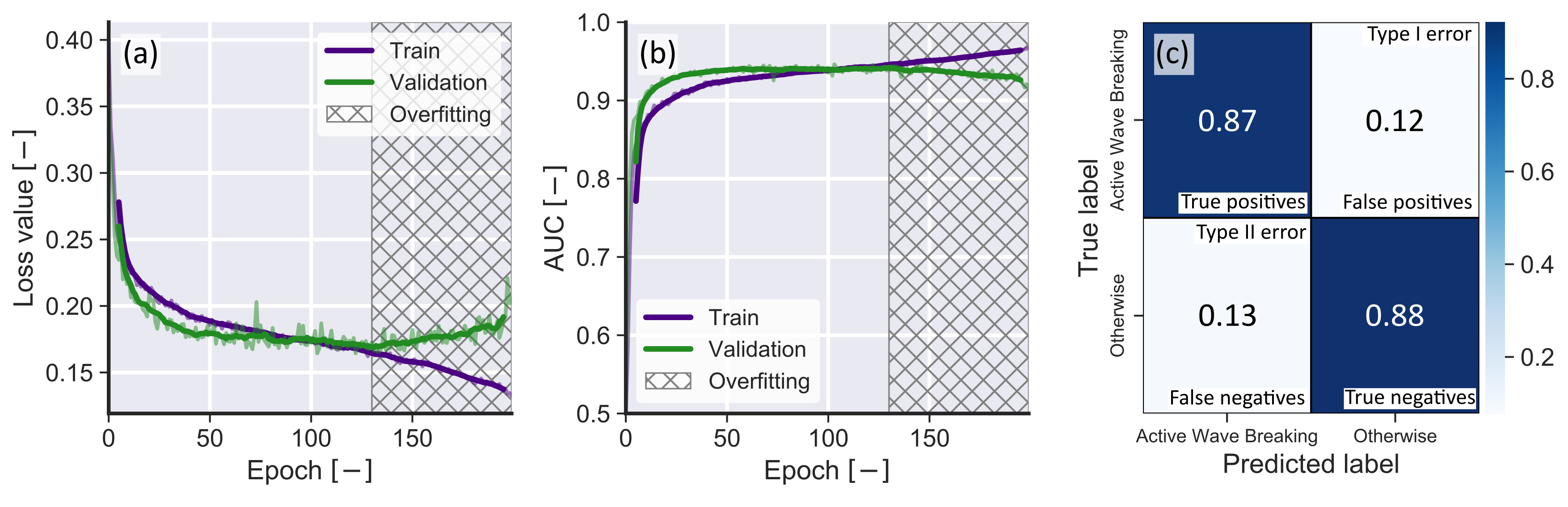}
    \caption{Examples of training curves and confusion matrix for the best overall performing model (VGG16). a) Loss function value for training and validation data. b) $AUC$ value for training and validation data. In a) and b) the hatched area indicates the epochs after which the model started to overfit, the thick colored lines show smoothed loss or $AUC$ values (average at every 10 epochs), and the transparent lines show raw loss or $AUC$ values. c) confusion matrix calculated using test data only.}
    \label{fig:training_and_cm}
\end{figure}

From the analysis of all training curves, confusion matrices, and from Table \ref{tab:classifier_results}, the best performing backbone architecture during training was ResNet50V2 by a considerable margin ($AUC$=0.989). These results however, did not translate to the validation dataset ($AUC$=0.873). Considering only the validation data, VGG16 was the best performing backbone with $AUC$=0.946. Considering only the test dataset, the best performing model was also VGG16 ($AUC$=0.855). Overall, VGG16 was selected as the best performing model and the results presented in the next sections will use this backbone. \add[Rev1]{Other evaluation metrics such as the accuracy score, precision, and recall closely followed the evolution of AUC with training epochs (compare Figures \ref{fig:training_and_cm}-a and b, for example). In general, as the loss value decreased, the number of false positives decreased, which made the precision and recall to increase. This behavior was consistent for all models.} \change[Rev1]{However, g}{G}iven that different architectures may perform better for other datasets and further \change[Rev1]{fine-tuning}{optimization} could change the model ranking presented here, all pre-trained models \add[Rev1]{and their training metrics evolution} are made available in \change[Rev1]{\url{https://github.com/caiostringari/deepwaves}}{the project's code repository}. Finally, it is worth mentioning that larger models (for example, VGG19, ResNET152 and EfficientNetB7) could achieve better results but this was not attempted there due to hardware limitations (that is, these models required more memory than what was available on the NVIDIA GTX 1080 GPU used in this study). 

\begin{table}[ht]
  \centering
  \caption{Evaluation metrics for all tested backbone architectures. Refer to the main manuscript text for definition of evaluation metrics. The best performing models are shown in boldface. Results are sorted by AUC.}
    \begin{tabular}{lrrrrrrrr}
    \textbf{Model} & \multicolumn{1}{l}{\textbf{Accuracy}} & \multicolumn{1}{l}{\textbf{TP}} & \multicolumn{1}{l}{\textbf{FP}} & \multicolumn{1}{l}{\textbf{TN}} & \multicolumn{1}{l}{\textbf{FN}} & \multicolumn{1}{l}{\textbf{Precision}} & \multicolumn{1}{l}{\textbf{Recall}} & \multicolumn{1}{l}{\textbf{AUC}} \\
    \midrule
    \multicolumn{9}{c}{\textbf{Train}} \\
    \midrule
    \textbf{ResNetV250} & \textbf{0.97} & \textbf{1414} & \textbf{198} & \textbf{13978} & \textbf{280} & \textbf{0.877} & \textbf{0.835} & \textbf{0.989} \\
    VGG16 & 0.93  & 855   & 273   & 13911 & 831   & 0.758 & 0.507 & 0.943 \\
    InceptionResnetV2 & 0.927 & 886   & 359   & 13823 & 802   & 0.712 & 0.525 & 0.932 \\
    EfficientNetB5 & 0.772 & 1403  & 3346  & 10920 & 297   & 0.295 & 0.825 & 0.874 \\
    MobileNet & 0.904 & 436   & 268   & 13916 & 1250  & 0.619 & 0.259 & 0.848 \\
    \midrule
    \multicolumn{9}{c}{\textbf{Validation}} \\
    \midrule
    \textbf{VGG16} & \textbf{0.932} & \textbf{221} & \textbf{65} & \textbf{3478} & \textbf{204} & \textbf{0.773} & \textbf{0.52} & \textbf{0.946} \\
    InceptionResnetV2 & 0.921 & 190   & 81    & 3466  & 231   & 0.701 & 0.451 & 0.93 \\
    EfficientNetB5 & 0.809 & 353   & 687   & 2856  & 72    & 0.339 & 0.831 & 0.897 \\
    MobileNet & 0.908 & 123   & 64    & 3479  & 302   & 0.658 & 0.289 & 0.878 \\
    ResNetV250 & 0.919 & 197   & 97    & 3450  & 224   & 0.67  & 0.468 & 0.873 \\
    \midrule
    \multicolumn{9}{c}{\textbf{Test}} \\
    \midrule
    \textbf{VGG16} & \textbf{0.876} & \textbf{106} & \textbf{80} & \textbf{945} & \textbf{69} & \textbf{0.57} & \textbf{0.606} & \textbf{0.855} \\
    ResNetV250 & 0.881 & 95    & 63    & 962   & 80    & 0.601 & 0.543 & 0.843 \\
    InceptionResnetV2 & 0.882 & 91    & 57    & 968   & 84    & 0.615 & 0.52  & 0.839 \\
    EfficientNetB5 & 0.873 & 88    & 65    & 960   & 87    & 0.575 & 0.503 & 0.827 \\
    MobileNet & 0.875 & 30    & 5     & 1020  & 145   & 0.857 & 0.171 & 0.768 \\
    \bottomrule
    \end{tabular}%
  \label{tab:classifier_results}%
\end{table}%

\subsection*{Test Case with Real-world Data}

\begin{figure}[htp]
    \centering
    \includegraphics[width=\linewidth]{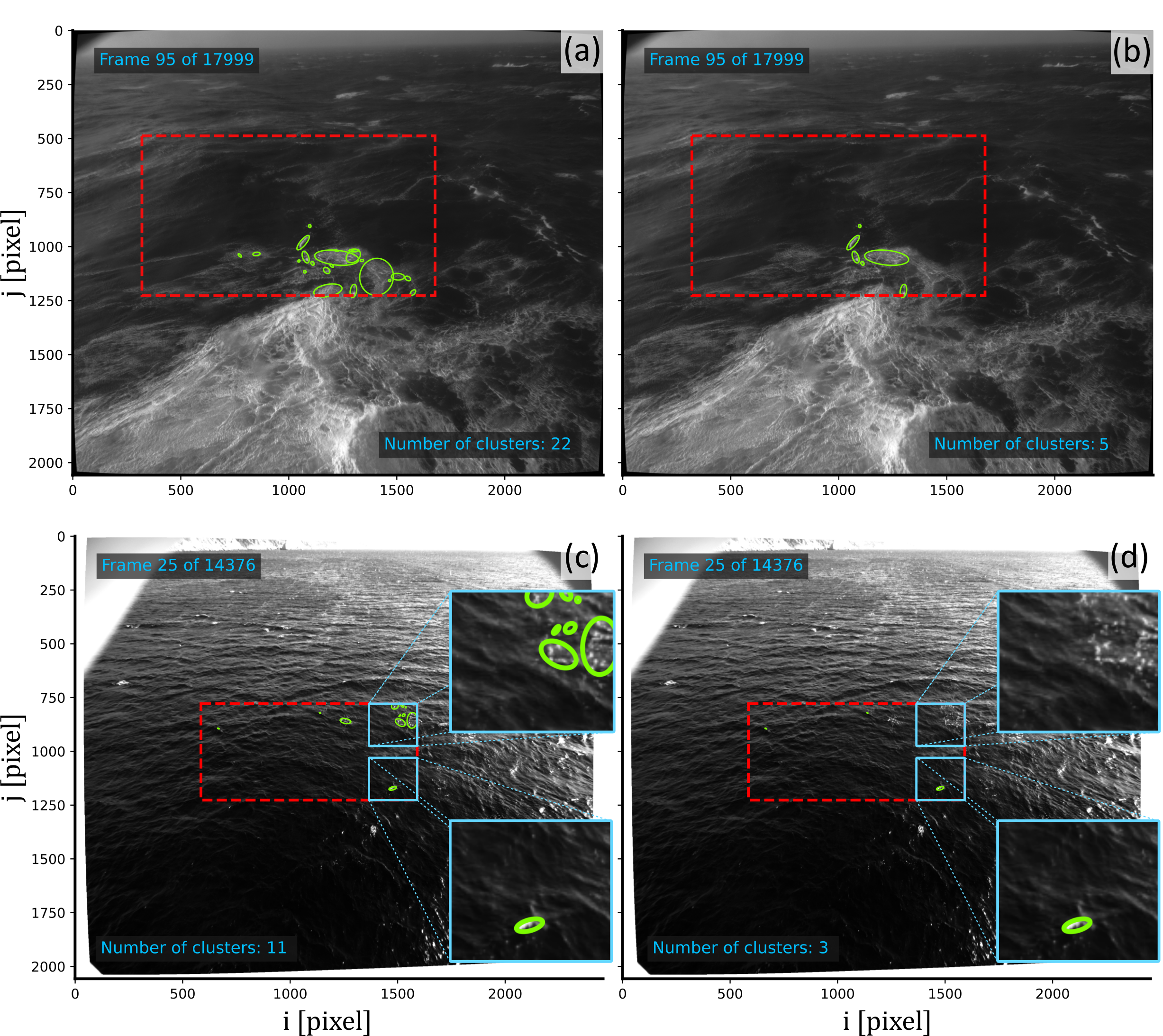}
    \caption{Example of the application of the method. a) Results of the na\"ive wave breaking detection (thresholding + DBSCAN) for La Jument data (03/01/2018 09:39). Note the great amount of passive foam being detected as active wave breaking. b) Results of active wave breaking detection using VGG16 as backbone for La Jument data (03/01/2018 09:39). Note the significant reduction in the amount of passive foam being detected. In both plots, number of clusters refers to the number of clusters identified by DBSCAN. The red dashed rectangle indicates the region of interest which, in these examples, was the same as the stereo-video reconstruction area. c) Results of the na\"ive wave breaking detection (thresholding + DBSCAN) for Black sea data (04/10/2011 09:38). d) Results of active wave breaking detection using VGG16 as backbone for Black Sea data (04/10/2011 09:38). Animations of these results are available at \url{https://github.com/caiostringari/deepwaves}. In this particular example, the image was subdivided into blocks of 256x256 pixels for processing. \add[Rev1]{Note that identical results were seen using other architectures other than VGG16 to classify these data.} }
    
    % \caption{Legend (350 words max). Example legend text.}
    \label{fig:robust_detector}
\end{figure}

Figure \ref{fig:robust_detector} shows the results of the application of the best performing model architecture (VGG16) on La Jument (2018/01/03 09:39) and Black Sea (2011/10/04 09:38) data. Visual inspection of these results confirmed the ability of the neural network to correctly classify the na\"ively identified wave breaking candidates and only keep instances that represented active wave breaking. Moreover, the same neural network was able to correctly classify active wave breaking events for the rogue waves seen at La Jument\cite{Filipot2019} and for the small wind-generated breakers seen in the Black Sea data. This result highlights the ability of the neural network to generalize well on the dataset, which is a difficult result to achieve. From the analysis of the training curves, the averaged classification error (accounting for the imbalance in the data) should be of the order of $\approx$10\%, which to the authors' knowledge it was not assessed by other wave breaking detection methods. We strongly recommend that future research should report active wave breaking detection errors when used to assess or further develop models that depend on wave breaking data.

\subsection*{Comparison with Mironov \& Dulov (2008)}
\begin{figure}[htp]
	\centering
	\includegraphics[width=0.8\textwidth]{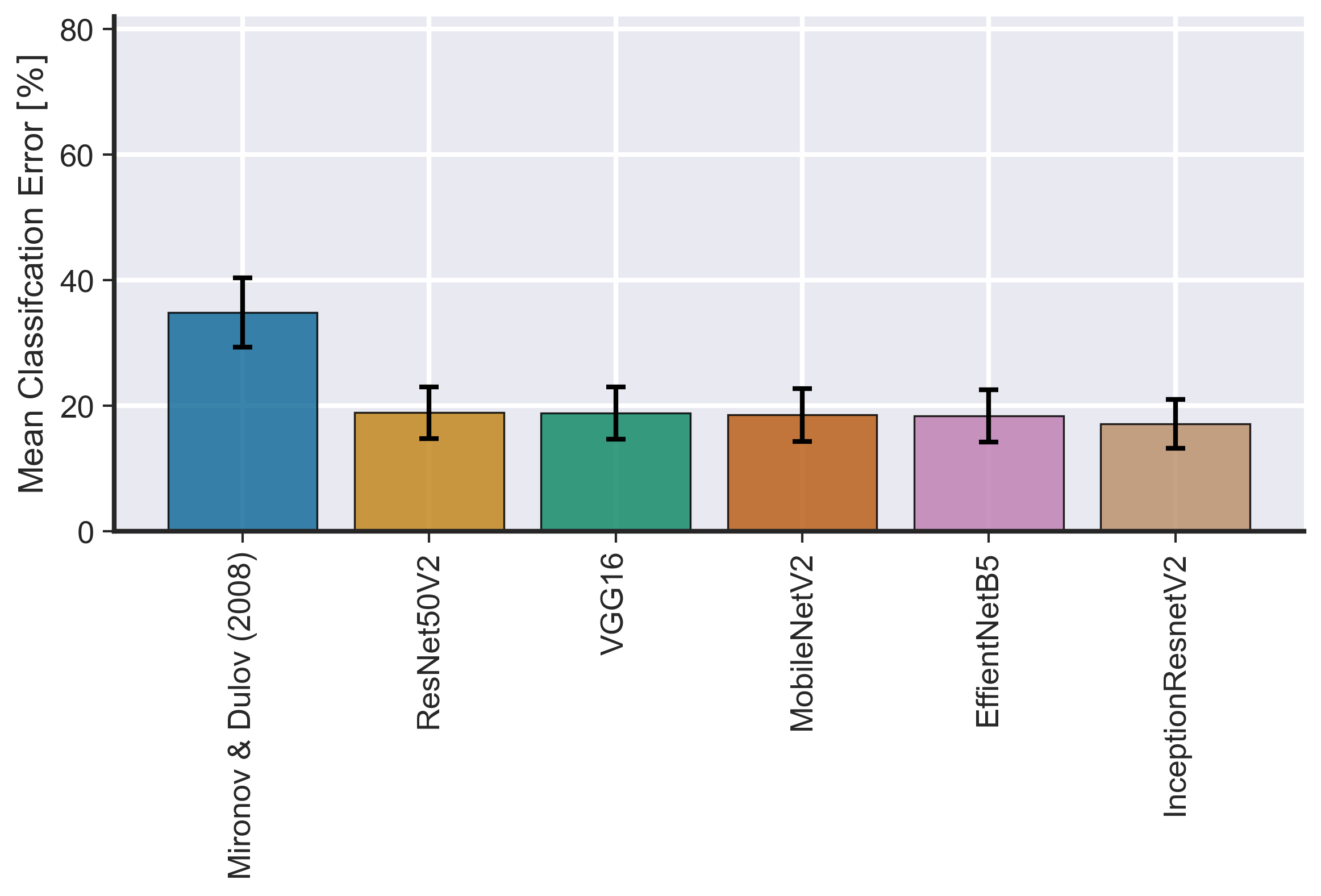}
	\caption{Comparison between averaged classification error for active wave breaking detection between Mironov \& Dulov's (2008)\cite{Mironov2008} method and all the neural network architectures implemented here. The error bars represent one standard deviation from the mean. The data used for this comparison are from the 2013 Black Sea experiments described in Table \ref{tab:dataset}.}\label{fig:comparison}
\end{figure}

\add[Rev1]{

To highlight the improvements obtained by the present method, this section presents a comparison between Mironov \& Dulov's (2008)\cite{Mironov2008} automatic active wave breaking detection method and the present method. To perform this task, data from the 2013 Black Sea experiments in Table \ref{tab:dataset} that had previously been classified and investigated in detail by Guimar\~aes\cite{Guimaraes2018} were used. All the active breaking events (that is, considering all 900s of data) detected using Mironov \& Dulov's (2008)\cite{Mironov2008} were manually classified as true or false and compared to the labels predicted by our method. To the authors' knowledge, these data are the only currently available data that has been classified by both methods as well as manually. Figure \ref{fig:comparison} shows the result of the compassion between models. On average, our method has relatively $\approx$50\% less error then Mironov \& Dulov's (2008) method with an averaged absolute reduction in error of $\approx$15\%. The results in Figure \ref{fig:comparison} are also consistent with the results seen in Table \ref{tab:classifier_results} which showed that our model had errors in the order of $\approx$15\% when considering the validation and test datasets. Note that all tested neural network architectures performed very similarly with only a slight advantage for InceptionResnetV2. It is also worth mentioning that Mironov \& Dulov's (2008) method was designed and optimized to work specifically with data from the Black Sea and it is not guaranteed that it will generalize to other datasets. Our method, on the contrary, has been shown to generalize well for very distinct datasets (see Figure \ref{fig:robust_detector}, for example) and with further optimization could achieve even better performance.}

\subsection*{Wave Breaking Statistics}

\begin{figure}[htp]
    \centering
    \includegraphics[width=\linewidth]{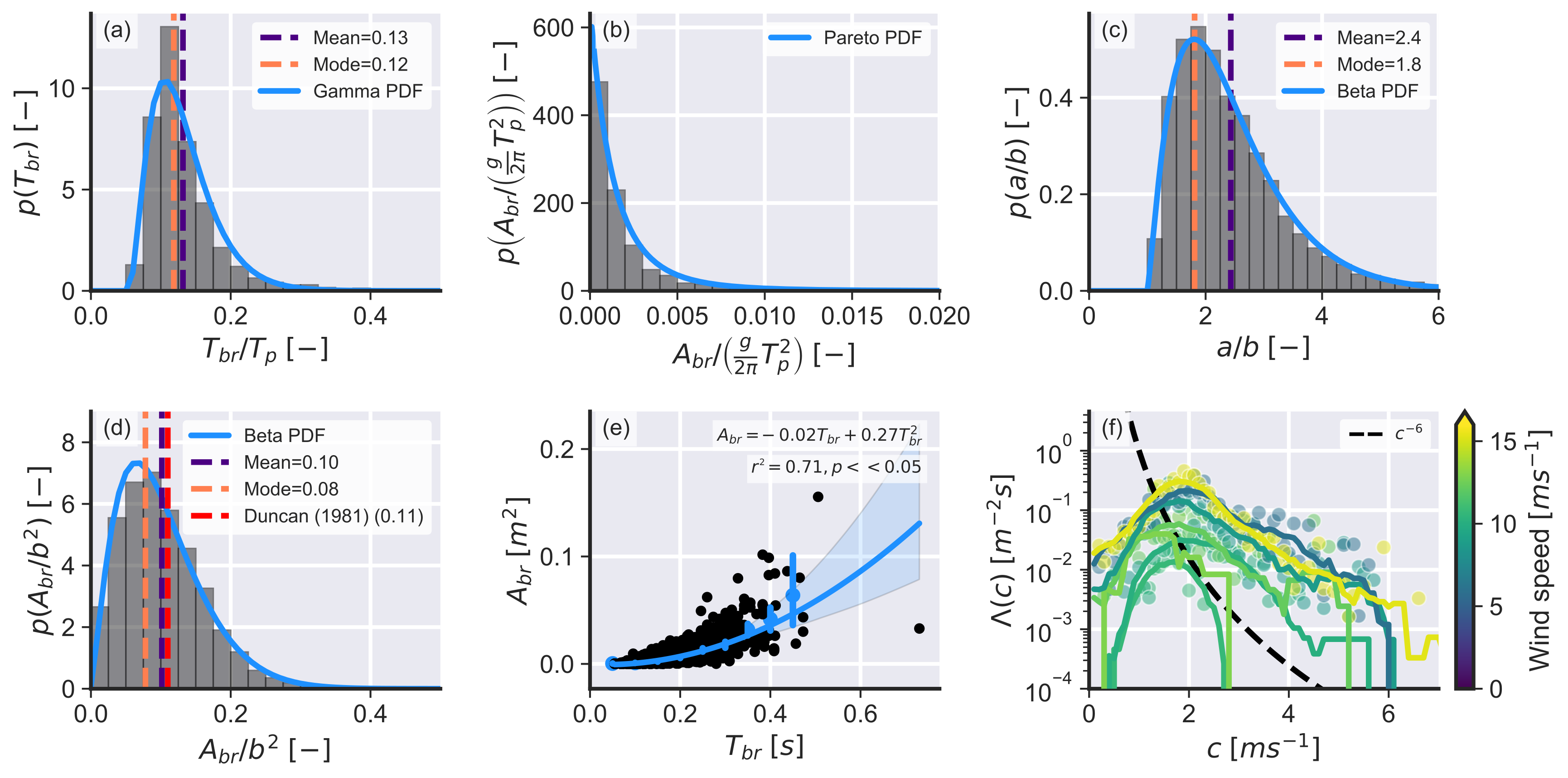}
    \caption{Examples of statistical properties of breaking waves that can be directly obtained from the proposed method. a) Probability distribution of the wave breaking duration ($T_{br}$) normalized by wave peak period ($T_p$). The blue line shows the Gamma fit to the data, the purple dashed line shows the mean value (0.13) and the orange dashed line shows the mode value (0.12). b) Probability distribution of the wave breaking area ($A_{br}$) normalized by wavelength ($\frac{g}{2\pi}T_p^2$). The blue line shows the Pareto fit to the data. c) Probability distribution of the ratio between major and minor axis ($a/b$). The blue line shows the Beta fit to the data, the purple dashed line shows the mean value (2.4) and the orange dashed line shows the mode value (1.8). d) Probability distribution of the wave breaking area scaling parameter ($A_{br}/b^2$). The blue line shows the Beta fit to the data, the purple dashed line shows the mean value (0.1), the orange dashed line shows the mode value (0.08), and the red line shows Duncan's constant value (0.11). e) Evolution of the wave breaking area over time. The black markers show the observed values, the error bars show values binned at 0.15s intervals, the blue line shows the quadratic regression to the data, the blue swath shows the 95\% confidence interval for the regression, and $r^2$ is the correlation coefficient.  f) Phillips\cite{Phillips1985} $\Lambda(c)dc$ distributions assuming that the wave breaking speed is the same as the phase speed of the carrier wave (that is, $c=c_{br}$) grouped by wind speed. The black dashed line shows the $c^{-6}$ theoretical decay, the colored markers show the average crest length binned at 0.1$ms^{-1}$ wave speed intervals, and the colored lines show a running average with a window size of 10 bins.}
    \label{fig:wave_stats}
\end{figure}

In this section we briefly present examples of wave breaking statistics that can be directly derived from the classified wave breaking data. For brevity, the analysis is limited to data from the Black Sea because our classifier performed best at this location (classification errors < 10\% \add[Rev1]{considering both 2011 and 2013 experiments}). Five quantities will be analysed: the wave breaking duration ($T_{br}$), the wave breaking area ($A_{br}$), the major ($a$) and minor ($b$) axis of the fitted ellipses (representative of the wave breaking lengths at their maximum during the active part of the wave breaking), and Phillips' distribution $\Lambda(c)dc$. These quantities were obtained directly from space-time clustered wave breaking events with the exception of the cumulative wave breaking area ($A_{br}$) which was calculated from the projections of pixels clustered in the first step of the method to metric coordinates. The results of this analyses are shown in Figure \ref{fig:wave_stats}.

The wave breaking duration ($T_{br}$ normalized by wave peak period ($T_p$), Figure \ref{fig:wave_stats}-a) roughly followed a shifted Gamma probability density function (PDF) and had a mean value of 0.12 and most frequent value (mode) of 0.13. This result shows that the active part of the wave breaking process happens very quickly. The wave breaking area ($A_{br}$, Figure \ref{fig:wave_stats}-b) closely followed a Pareto distribution which indicates that large wave breaking events are relatively rare in the data. The ratio between the major and minor axis of the fitted ellipses ($a/b$, Figure \ref{fig:wave_stats}-c) followed a Beta PDF and had a mean of 2.5 and mode of 1.9, which indicates that the ellipses' major axis is approximately double the size of the minor axis. Assuming a negligible wave front angle, the wave breaking area scaling relation from Duncan \cite{Duncan1981} ($A_{br}/b^{2}$, Figure \ref{fig:wave_stats}-d) also followed a Beta PDF and had mean of 0.1 and mode of 0.8, which is consistent with the previously reported value\cite{Duncan1981} (0.1$\pm$0.01). The wave breaking area (Figure \ref{fig:wave_stats}-e) showed a quadratic increase with wave breaking event duration, which is trivial but seems not to have been previously directly shown before. Finally, Figure \ref{fig:wave_stats}-f shows the $\Lambda(c)dc$ distributions obtained using our method and considering that the ellipse major axis is representative of the wave breaking length. The observed distributions greatly deviated from the theoretical $c^{-6}$ relation\cite{Phillips1985}. Note that all PDF fits shown here were statistically significant at the 95\% confidence level using the two-tailed Kolmogorov–Smirnov test. See Table \ref{tab:pdfs} for the description of the parameters of the PDFs presented in this section and the Discussion section for the contextualization of these results and the possible implications that they have for future research.

\begin{table}[ht]
  \centering
  \caption{Fitted parameters for the distributions seen in Figure \ref{fig:wave_stats} . All fits were done using \textit{Scipy}\cite{2020SciPy-NMeth} and, consequently, the values reported here follow \textit{Scipy}'s conventions.}
    \begin{tabular}{l l l l l l l l}
    \toprule
      & \textbf{Distribution} & \textbf{Parameter 1}  & \textbf{Parameter 2} & \textbf{Location} & \textbf{Scale} \\
    \midrule
    Figure \ref{fig:wave_stats}-a &  Gamma & 3.001 & - & 0.053 & 0.0260 \\
    Figure \ref{fig:wave_stats}-b &  Pareto  & 4.339 & - & -0.007 &  0.007  \\
    Figure \ref{fig:wave_stats}-c &  Beta & 6.117 & 73.948 & 0.207 & 30.440 \\
    Figure \ref{fig:wave_stats}-d &  Beta  & 6.389 & 40.470 & -0.760& 24.438 \\
    \bottomrule
    \end{tabular}%
  \label{tab:pdfs}%
\end{table}%

\section*{Discussion}

We have presented a new method to detect active wave breaking in video imagery data that is  robust and easily transferable to future studies. Overall, VGG16 was the best performing architecture which is a surprising result given that VGG is a considerably older architecture that has been superseded by more recent architectures such as ResNets and EfficientNets \cite{Tan2019}. Also surprisingly, EfficientNet was one of the worst performers \add[Rev1]{considering the test dataset} despite the state-of-the-art results that this architecture achieved in recent years\cite{Tan2019}. One explanation could be that given VGG16 is the model with the highest number of parameters \add[Rev1]{(despite being the shallowest network)}, it adapted better to the characteristics of the present dataset. Another explanation could be that EfficientNet is currently overfit on the ImageNet\cite{Russakovsky2015} dataset that was used to train this network hence its high reported classification scores. Another aspect to take into consideration is the speed in which the models can predict on new data (inference). While VGG16 was the best performing model, its size slows down inference time. In this regard, MobileNetV2 offers real-time inference speed at 10Hz \add[Rev1]{for small image sizes (for example $512 \times 512$px)}. Consequently, a model based on MobileNetV2 could be used to detect active wave breaking in an operational fashion in, for example, offshore floating wind turbines that are susceptible to wave breaking impacts and used to adjust anchor and cable settings in real-time.

From the analysis of the application of the method on real-world data, it was visually observed that small active wave breaking events were not always detected, particularly for the Black Sea data. There are two possible explanations for this error. The simplest explanation could be that this type of events is under-represented in the training dataset. The solution for this issue consists of adding more data representative of these instances to the training dataset. The second possibility is that the image size becomes too small in the deeper layers of the network which makes it impossible for the network to learn these events (see below for further discussion). A solution for this issue could be to increase the input image size but this was not attempted here due to hardware constraints (that is, memory limitation, as discussed above).

\begin{figure}[ht]
    \centering
    \includegraphics[width=\linewidth]{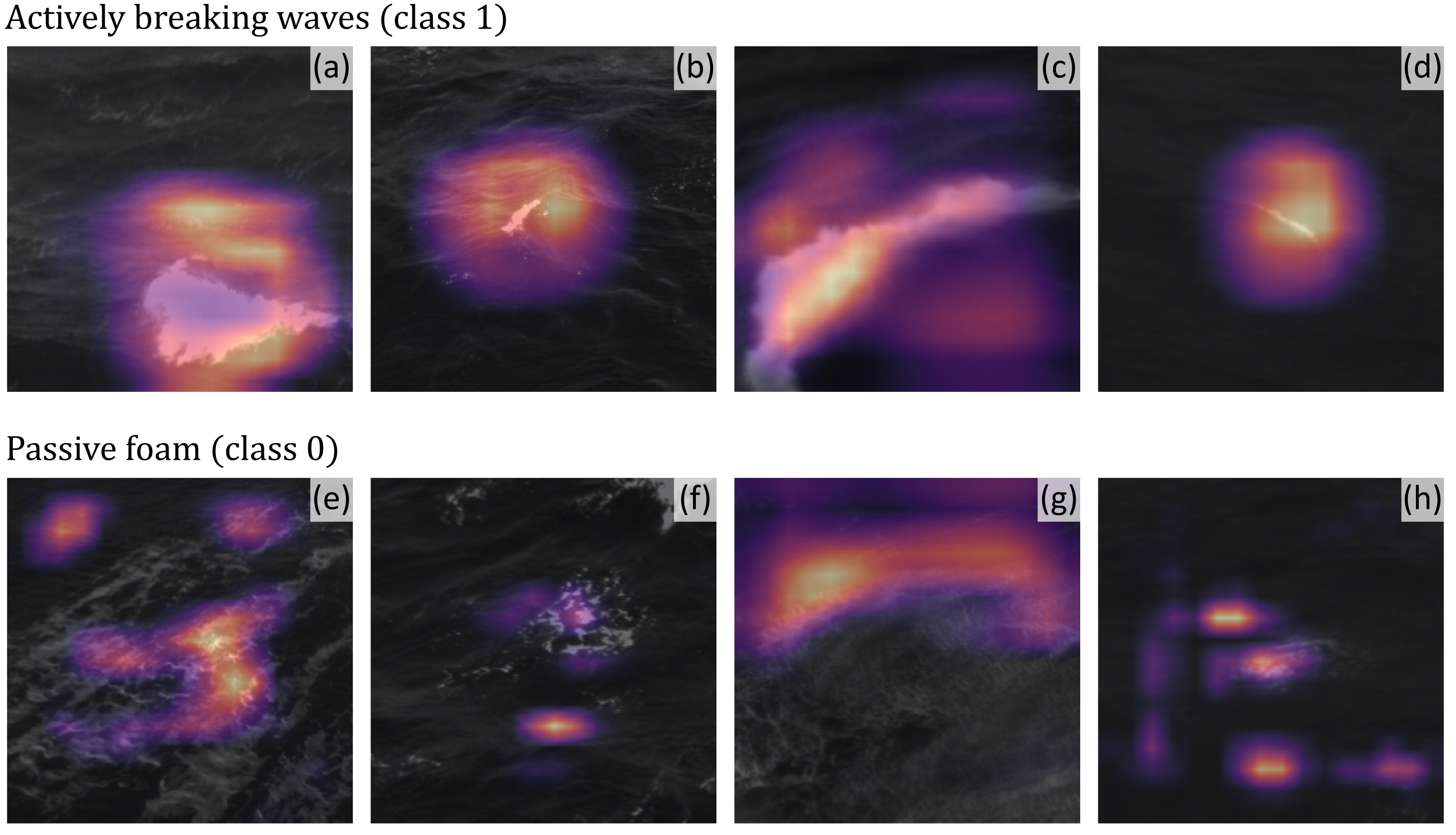}
    \caption{Results of Grad-CAM \cite{Selvaraju2020} for all unique experiment locations described in Table \ref{tab:dataset} applied to actively breaking waves (top row) and to passive foam (bottom row). a) Actively breaking wave example recorded at Adriatic Sea (2015/03/05 10:35). b) Actively breaking wave example recorded at the Black Sea (2011/10/04 11:07). c) Actively breaking wave example recorded at La Jument (2018/01/03 09:39). d) Actively breaking wave example recorded at the Yellow Sea (2017/05/13 05:00). e) Passive foam example recorded at Adriatic Sea (2015/03/05 10:35). f) Passive foam example recorded at the Black Sea (2011/10/04 11:07). g) Passive foam example recorded at La Jument (2018/01/03 09:39). h) Passive foam example recorded at the (2017/05/13 05:00). In all panels, the color scale indicates the weights of the class activation map with brighter colors showing regions of the image which the neural network used to classify each particular sample. }
    \label{fig:gradcam}
\end{figure}

Neural networks have been historically seen as black boxes in which only the final classification outputs are relevant. There has been, however, an increase in interest to understanding how neural networks are learning. One technique that is of particular interest for the present paper is the concept of Gradient-weighted Class Activation Mapping (Grad-CAM)\cite{Selvaraju2020}. Briefly, this technique shows which regions of a convolutional layer are more important for the results obtained by the classifier. Figure \ref{fig:gradcam} shows the results of Grad-CAM applied to examples for all unique locations from Table \ref{tab:dataset} considering VGG16's last convolutional layer. When considering only actively breaking waves (Figure \ref{fig:gradcam}-a to d) it is evident that VGG16 closely mimicked how a human would classify these data, that is, it directly searched for the regions of the image that corresponded to active wave breaking. In the case of passive foam (Figure \ref{fig:gradcam}-e to h), VGG16 seemed to use larger portions of the image but at the same time focused on the flocculent foam seen in the images as a human classifier would do. In general, these results show that our model truly learned how to classify the images and is not merely guessing the labels.

The promising results presented here indicate that the current method should be extended to an object detection framework. Such a framework would eliminate the need for the image thresholding and DBSCAN steps. This implementation could be done by using a strongly-supervised  architecture such as UNet \cite{RFB15a}, or by using a weakly-supervised method derived from Grad-CAM, for example. As a final recommendation regarding the machine learning aspect of this paper, we strongly encourage future researchers to add samples to the training dataset that matches their specific research needs instead of blindly applying the provided pre-trained models. 

We have also presented examples of wave breaking statistics that can be obtained using the proposed method. In general, the patterns observed here agreed with previously reported scaling factors that support the idea of wave breaking self-similarity. For example, Figure \ref{fig:wave_stats}-f directly showed that the scaling parameter $A_{br}/b^2$ approaches the constant 0.1 value from Duncan's \cite{Duncan1981} laboratory experiments. Another variable that showed signs of a self-similar behavior was the wave breaking area ($A_{br}$) which was very well described by a Pareto distribution and presented a steady quadratic increase with wave breaking duration ($T_{br}$). Extensive research \cite{ Melville2002, Romero2019} has been grounded on the assumption that wave breaking is self-similar, but inconsistencies with this approach have been reported before \cite{Gemmrich2008}. Contrarily to other studies \cite{Melville2002, Sutherland2013}, however, the $\Lambda(c)$ distributions obtained here did not closely match the theoretical $c^{-6}$ for a sea-state in equilibrium. As reported before \cite{Banner2014}, these differences may be due to the fact that here we only considered actively breaking waves for our analysis whereas other studies seem to track the speeds of both actively breaking waves and passive foam \cite{Kleiss2010, Sutherland2013}. \add[Rev1]{Another possibility is that other phenomena not accounted by Phillips'\cite{Phillips1985} theory (for example, wave modulation\cite{Donelan2010}) play important role on wave breaking.} A future publication that \change[Rev1]{more deeply explores}{investigates the mechanisms related to} the wave breaking kinematics \add[Rev1]{using the method and} data obtained here will soon follow.

\section*{Conclusion}

We described a novel method to detect and classify actively breaking waves in video imagery data. Our method achieved promising results when assessed using several different metrics. Further, by analyzing the deeper layers of our neural network, we showed that the model mimicked how a human classifier would perform a similar classification task. As an application of the method, we presented wave breaking statistics and breaking wave crest length distributions. Our method can thus be useful for the investigation of several ocean-atmosphere interaction processes and should, in the future, lead to advancements in operational wave forecast models, gas-exchange models, and general safety at seas. Finally, we strongly recommend that future research should focus on standardized methods to study wave breaking so that a consistent dataset can be generated and used freely and unambiguously by the community.  

\bibliography{library}

\begin{thebibliography}{10}
\urlstyle{rm}
\expandafter\ifx\csname url\endcsname\relax
  \def\url#1{\texttt{#1}}\fi
\expandafter\ifx\csname urlprefix\endcsname\relax\def\urlprefix{URL }\fi
\expandafter\ifx\csname doiprefix\endcsname\relax\def\doiprefix{DOI: }\fi
\providecommand{\bibinfo}[2]{#2}
\providecommand{\eprint}[2][]{\url{#2}}

\bibitem{Battjes1978}
\bibinfo{author}{Battjes, J.~A.} \& \bibinfo{author}{Janssen, J.}
\newblock \bibinfo{journal}{\bibinfo{title}{{Energy loss and set-up due to
  breaking of random waves}}}.
\newblock {\emph{\JournalTitle{Coastal Engineering}}}
  \textbf{\bibinfo{volume}{32}}, \bibinfo{pages}{569--587}
  (\bibinfo{year}{1978}).

\bibitem{Thornton1983}
\bibinfo{author}{Thornton, E.~B.} \& \bibinfo{author}{Guza, R.~T.}
\newblock \bibinfo{journal}{\bibinfo{title}{{Transformation of Wave Height
  Distribution}}}.
\newblock {\emph{\JournalTitle{Journal of Geophysical Research}}}
  \textbf{\bibinfo{volume}{88}}, \bibinfo{pages}{5925--5938}
  (\bibinfo{year}{1983}).

\bibitem{Banner2000}
\bibinfo{author}{Banner, M.~L.}, \bibinfo{author}{Babanin, A.~V.} \&
  \bibinfo{author}{Young, I.~R.}
\newblock \bibinfo{journal}{\bibinfo{title}{{Breaking Probability for Dominant
  Waves on the Sea Surface}}}.
\newblock {\emph{\JournalTitle{Journal of Physical Oceanography}}}
  \textbf{\bibinfo{volume}{30}}, \bibinfo{pages}{3145--3160},
  \doiprefix\url{10.1175/1520-0485(2000)030<3145:BPFDWO>2.0.CO;2}
  (\bibinfo{year}{2000}).

\bibitem{Banner2002}
\bibinfo{author}{Banner, M.~L.}, \bibinfo{author}{Gemmrich, J.~R.} \&
  \bibinfo{author}{Farmer, D.~M.}
\newblock \bibinfo{journal}{\bibinfo{title}{{Multiscale measurements of ocean
  wave breaking probability}}}.
\newblock {\emph{\JournalTitle{Journal of Physical Oceanography}}}
  \textbf{\bibinfo{volume}{32}}, \bibinfo{pages}{3364--3375},
  \doiprefix\url{10.1175/1520-0485(2002)032<3364:MMOOWB>2.0.CO;2}
  (\bibinfo{year}{2002}).

\bibitem{Cavaleri2006}
\bibinfo{author}{Cavaleri, B. Y.~L.}
\newblock \bibinfo{journal}{\bibinfo{title}{{Wave Modeling: Where to Go in the
  Future}}}.
\newblock {\emph{\JournalTitle{Bulletin of American Meteorological Society}}}
  \bibinfo{pages}{207--2014}, \doiprefix\url{10.1175/BAMS-87-2-207}
  (\bibinfo{year}{2006}).

\bibitem{WW32019}
\bibinfo{author}{{The WAVEWATCH Development Group (WW3DG)}}.
\newblock \bibinfo{title}{{User manual and system documentation of WAVEWATCH
  III R version 6.07}}.
\newblock \bibinfo{type}{Tech. Rep.},
  \bibinfo{institution}{NOAA/NWS/NCEP/MMAB}, \bibinfo{address}{College Park,
  MD, USA} (\bibinfo{year}{2019}).

\bibitem{Booij1999}
\bibinfo{author}{Booij, N.}, \bibinfo{author}{Ris, R.~C.} \&
  \bibinfo{author}{Holthuijsen, L.~H.}
\newblock \bibinfo{journal}{\bibinfo{title}{{A third-generation wave model for
  coastal regions 1 . Model description and validation}}}.
\newblock {\emph{\JournalTitle{Journal of Geophysical Research}}}
  \textbf{\bibinfo{volume}{104}}, \bibinfo{pages}{7649--7666}
  (\bibinfo{year}{1999}).

\bibitem{THEWANDIGROUP1988}
\bibinfo{author}{{The WANDI Group}}.
\newblock \bibinfo{journal}{\bibinfo{title}{{The WAN Model - A Third Generation
  Ocean Wave Prediction Model}}}.
\newblock {\emph{\JournalTitle{Journal of Physical Oceanography}}}
  \textbf{\bibinfo{volume}{18}}, \bibinfo{pages}{1775--1810}
  (\bibinfo{year}{1988}).

\bibitem{Filipot2010a}
\bibinfo{author}{Filipot, J.~F.}, \bibinfo{author}{Ardhuin, F.} \&
  \bibinfo{author}{Babanin, A.~V.}
\newblock \bibinfo{journal}{\bibinfo{title}{{A unified deep-to-shallow water
  wave-breaking probability parameterization}}}.
\newblock {\emph{\JournalTitle{Journal of Geophysical Research: Oceans}}}
  \textbf{\bibinfo{volume}{115}}, \bibinfo{pages}{1--15},
  \doiprefix\url{10.1029/2009JC005448} (\bibinfo{year}{2010}).

\bibitem{Stringari2019a}
\bibinfo{author}{Stringari, C.~E.} \& \bibinfo{author}{Power, H.~E.}
\newblock \bibinfo{journal}{\bibinfo{title}{{The Fraction of Broken Waves in
  Natural Surf Zones}}}.
\newblock {\emph{\JournalTitle{Journal of Geophysical Research: Oceans}}}
  \textbf{\bibinfo{volume}{124}}, \bibinfo{pages}{1--27},
  \doiprefix\url{10.1029/2019JC015213} (\bibinfo{year}{2019}).
\newblock \eprint{arXiv:1904.06821v1}.

\bibitem{Melville2002}
\bibinfo{author}{Melville, W.~K.} \& \bibinfo{author}{Matusov, P.}
\newblock \bibinfo{journal}{\bibinfo{title}{{Distribution of breaking waves at
  the ocean surface}}}.
\newblock {\emph{\JournalTitle{Nature}}} \textbf{\bibinfo{volume}{417}},
  \bibinfo{pages}{58--63}, \doiprefix\url{10.1038/417058a}
  (\bibinfo{year}{2002}).

\bibitem{Phillips1985}
\bibinfo{author}{Phillips, O.~M.}
\newblock \bibinfo{journal}{\bibinfo{title}{{Spectral and statistical
  properties of the equilibrium range in wind-generated gravity waves}}}.
\newblock {\emph{\JournalTitle{Journal of Fluid Mechanics}}}
  \textbf{\bibinfo{volume}{156}}, \bibinfo{pages}{505--531},
  \doiprefix\url{10.1017/S0022112085002221} (\bibinfo{year}{1985}).

\bibitem{Banner2014}
\bibinfo{author}{Banner, M.~L.}, \bibinfo{author}{Zappa, C.~J.} \&
  \bibinfo{author}{Gemmrich, J.~R.}
\newblock \bibinfo{journal}{\bibinfo{title}{{A note on the Phillips spectral
  framework for ocean whitecaps}}}.
\newblock {\emph{\JournalTitle{Journal of Physical Oceanography}}}
  \textbf{\bibinfo{volume}{44}}, \bibinfo{pages}{1727--1734},
  \doiprefix\url{10.1175/JPO-D-13-0126.1} (\bibinfo{year}{2014}).

\bibitem{Gemmrich2008}
\bibinfo{author}{Gemmrich, J.~R.}, \bibinfo{author}{Banner, M.~L.} \&
  \bibinfo{author}{Garrett, C.}
\newblock \bibinfo{journal}{\bibinfo{title}{{Spectrally resolved energy
  dissipation rate and momentum flux of breaking waves}}}.
\newblock {\emph{\JournalTitle{Journal of Physical Oceanography}}}
  \textbf{\bibinfo{volume}{38}}, \bibinfo{pages}{1296--1312},
  \doiprefix\url{10.1175/2007JPO3762.1} (\bibinfo{year}{2008}).

\bibitem{Romero2019}
\bibinfo{author}{Romero, L.}
\newblock \bibinfo{journal}{\bibinfo{title}{{Distribution of Surface Wave
  Breaking Fronts}}}.
\newblock {\emph{\JournalTitle{Geophysical Research Letters}}}
  \textbf{\bibinfo{volume}{46}}, \bibinfo{pages}{10463--10474},
  \doiprefix\url{10.1029/2019GL083408} (\bibinfo{year}{2019}).

\bibitem{Yurovsky2018}
\bibinfo{author}{Yurovsky, Y.~Y.}, \bibinfo{author}{Kudryavtsev, V.~N.},
  \bibinfo{author}{Chapron, B.} \& \bibinfo{author}{Grodsky, S.~A.}
\newblock \bibinfo{journal}{\bibinfo{title}{{Modulation of Ka-Band Doppler
  Radar Signals Backscattered from the Sea Surface}}}.
\newblock {\emph{\JournalTitle{IEEE Transactions on Geoscience and Remote
  Sensing}}} \textbf{\bibinfo{volume}{56}}, \bibinfo{pages}{2931--2948},
  \doiprefix\url{10.1109/TGRS.2017.2787459} (\bibinfo{year}{2018}).

\bibitem{Huang2017}
\bibinfo{author}{Huang, L.}, \bibinfo{author}{Liu, B.}, \bibinfo{author}{Li,
  X.}, \bibinfo{author}{Zhang, Z.} \& \bibinfo{author}{Yu, W.}
\newblock \bibinfo{journal}{\bibinfo{title}{{Technical evaluation of Sentinel-1
  IW mode cross-pol radar backscattering from the ocean surface in moderate
  wind condition}}}.
\newblock {\emph{\JournalTitle{Remote Sensing}}} \textbf{\bibinfo{volume}{9}},
  \bibinfo{pages}{1--21}, \doiprefix\url{10.3390/rs9080854}
  (\bibinfo{year}{2017}).

\bibitem{Hwang2010}
\bibinfo{author}{Hwang, P.~A.}, \bibinfo{author}{Zhang, B.},
  \bibinfo{author}{Toporkov, J.~V.} \& \bibinfo{author}{Perrie, W.}
\newblock \bibinfo{journal}{\bibinfo{title}{{Comparison of composite Bragg
  theory and quad-polarization radar backscatter from RADARSAT-2: With
  applications to wave breaking and high wind retrieval}}}.
\newblock {\emph{\JournalTitle{Journal of Geophysical Research: Oceans}}}
  \textbf{\bibinfo{volume}{115}}, \bibinfo{pages}{1--12},
  \doiprefix\url{10.1029/2009JC005995} (\bibinfo{year}{2010}).

\bibitem{Monahan2002}
\bibinfo{author}{Monahan, E.~C.}
\newblock \bibinfo{journal}{\bibinfo{title}{Oceanic whitecaps: Sea surface
  features detectable via satellite that are indicators of the magnitude of the
  air-sea gas transfer coefficient}}.
\newblock {\emph{\JournalTitle{Journal of Earth System Science}}}
  \textbf{\bibinfo{volume}{111}}, \doiprefix\url{10.1007/BF02701977}
  (\bibinfo{year}{2002}).

\bibitem{Reul2003}
\bibinfo{author}{Reul, N.} \& \bibinfo{author}{Chapron, B.}
\newblock \bibinfo{journal}{\bibinfo{title}{{A model of sea-foam thickness
  distribution for passive microwave remote sensing applications}}}.
\newblock {\emph{\JournalTitle{Journal of Geophysical Research C: Oceans}}}
  \textbf{\bibinfo{volume}{108}}, \bibinfo{pages}{19--1},
  \doiprefix\url{10.1029/2003jc001887} (\bibinfo{year}{2003}).

\bibitem{Carini2015}
\bibinfo{author}{Carini, R.~J.}, \bibinfo{author}{Chickadel, C.~C.},
  \bibinfo{author}{Jessup, A.~T.} \& \bibinfo{author}{Thompson, J.}
\newblock \bibinfo{journal}{\bibinfo{title}{{Estimating wave energy dissipation
  in the surf zone using thermal infrared imagery}}}.
\newblock {\emph{\JournalTitle{Journal of Geophysical Research: Oceans}}}
  \textbf{\bibinfo{volume}{120}}, \bibinfo{pages}{3937--3957},
  \doiprefix\url{10.1002/2014JC010561.Received} (\bibinfo{year}{2015}).

\bibitem{Wang2019}
\bibinfo{author}{Wang, S.} \emph{et~al.}
\newblock \bibinfo{journal}{\bibinfo{title}{{Improving the Upper-Ocean
  Temperature in an Ocean Climate Model (FESOM 1.4): Shortwave Penetration
  Versus Mixing Induced by Nonbreaking Surface Waves}}}.
\newblock {\emph{\JournalTitle{Journal of Advances in Modeling Earth Systems}}}
  \textbf{\bibinfo{volume}{11}}, \bibinfo{pages}{545--557},
  \doiprefix\url{10.1029/2018MS001494} (\bibinfo{year}{2019}).

\bibitem{Komori2018}
\bibinfo{author}{Komori, S.} \emph{et~al.}
\newblock \bibinfo{journal}{\bibinfo{title}{{Laboratory measurements of heat
  transfer and drag coefficients at extremely high wind speeds}}}.
\newblock {\emph{\JournalTitle{Journal of Physical Oceanography}}}
  \textbf{\bibinfo{volume}{48}}, \bibinfo{pages}{959--974},
  \doiprefix\url{10.1175/JPO-D-17-0243.1} (\bibinfo{year}{2018}).

\bibitem{Buscombe2019}
\bibinfo{author}{Buscombe, D.} \& \bibinfo{author}{Carini, R.~J.}
\newblock \bibinfo{journal}{\bibinfo{title}{{A data-driven approach to
  classifying wave breaking in infrared imagery}}}.
\newblock {\emph{\JournalTitle{Remote Sensing}}} \textbf{\bibinfo{volume}{11}},
  \bibinfo{pages}{1--10}, \doiprefix\url{10.3390/RS11070859}
  (\bibinfo{year}{2019}).

\bibitem{Buscombe2020}
\bibinfo{author}{Buscombe, D.}, \bibinfo{author}{Carini, R.~J.},
  \bibinfo{author}{Harrison, S.~R.}, \bibinfo{author}{Chickadel, C.~C.} \&
  \bibinfo{author}{Warrick, J.~A.}
\newblock \bibinfo{journal}{\bibinfo{title}{{Optical wave gauging using deep
  neural networks}}}.
\newblock {\emph{\JournalTitle{Coastal Engineering}}}
  \textbf{\bibinfo{volume}{155}}, \bibinfo{pages}{103593},
  \doiprefix\url{10.1016/j.coastaleng.2019.103593} (\bibinfo{year}{2020}).

\bibitem{Kim2020}
\bibinfo{author}{Kim, J.}, \bibinfo{author}{Kim, J.}, \bibinfo{author}{Kim,
  T.}, \bibinfo{author}{Huh, D.} \& \bibinfo{author}{Caires, S.}
\newblock \bibinfo{journal}{\bibinfo{title}{{Wave-tracking in the surf zone
  using coastal video imagery with deep neural networks}}}.
\newblock {\emph{\JournalTitle{Atmosphere}}} \textbf{\bibinfo{volume}{11}},
  \bibinfo{pages}{1--13}, \doiprefix\url{10.3390/atmos11030304}
  (\bibinfo{year}{2020}).

\bibitem{Bieman2020}
\bibinfo{author}{Bieman, J. P.~D.}, \bibinfo{author}{Ridder, M. P.~D.} \&
  \bibinfo{author}{Gent, M. R. A.~V.}
\newblock \bibinfo{journal}{\bibinfo{title}{{Deep learning video analysis as
  measurement technique in physical models}}}.
\newblock {\emph{\JournalTitle{Coastal Engineering}}}
  \textbf{\bibinfo{volume}{158}}, \bibinfo{pages}{103689},
  \doiprefix\url{10.1016/j.coastaleng.2020.103689} (\bibinfo{year}{2020}).

\bibitem{Zheng2019}
\bibinfo{author}{Zheng, C.~W.}, \bibinfo{author}{Chen, Y.~G.},
  \bibinfo{author}{Zhan, C.} \& \bibinfo{author}{Wang, Q.}
\newblock \bibinfo{journal}{\bibinfo{title}{{Source Tracing of the Swell
  Energy: A Case Study of the Pacific Ocean}}}.
\newblock {\emph{\JournalTitle{IEEE Access}}} \textbf{\bibinfo{volume}{7}},
  \bibinfo{pages}{139264--139275}, \doiprefix\url{10.1109/ACCESS.2019.2943903}
  (\bibinfo{year}{2019}).

\bibitem{Filipot2019}
\bibinfo{author}{Filipot, J.-F.} \emph{et~al.}
\newblock \bibinfo{journal}{\bibinfo{title}{{La Jument lighthouse: a real-scale
  laboratory for the study of giant waves and their loading on marine
  structures}}}.
\newblock {\emph{\JournalTitle{Philosophical Transactions of the Royal Society
  A: Mathematical, Physical and Engineering Sciences}}}
  \textbf{\bibinfo{volume}{377}}, \bibinfo{pages}{20190008},
  \doiprefix\url{10.1098/rsta.2019.0008} (\bibinfo{year}{2019}).

\bibitem{Mironov2008}
\bibinfo{author}{Mironov, A.~S.} \& \bibinfo{author}{Dulov, V.~A.}
\newblock \bibinfo{journal}{\bibinfo{title}{{Detection of wave breaking using
  sea surface video records}}}.
\newblock {\emph{\JournalTitle{Measurement Science and Technology}}}
  \textbf{\bibinfo{volume}{19}}, \doiprefix\url{10.1088/0957-0233/19/1/015405}
  (\bibinfo{year}{2008}).

\bibitem{Sutherland2013}
\bibinfo{author}{Sutherland, P.} \& \bibinfo{author}{Melville, W.~K.}
\newblock \bibinfo{journal}{\bibinfo{title}{{Field measurements and scaling of
  ocean surface wave-breaking statistics}}}.
\newblock {\emph{\JournalTitle{Geophysical Research Letters}}}
  \textbf{\bibinfo{volume}{40}}, \bibinfo{pages}{3074--3079},
  \doiprefix\url{10.1002/grl.50584} (\bibinfo{year}{2013}).

\bibitem{Kleiss2010}
\bibinfo{author}{Kleiss, J.~M.} \& \bibinfo{author}{Melville, W.~K.}
\newblock \bibinfo{journal}{\bibinfo{title}{{Observations of wave breaking
  kinematics in fetch-limited seas}}}.
\newblock {\emph{\JournalTitle{Journal of Physical Oceanography}}}
  \textbf{\bibinfo{volume}{40}}, \bibinfo{pages}{2575--2604},
  \doiprefix\url{10.1175/2010JPO4383.1} (\bibinfo{year}{2010}).

\bibitem{OpenCV}
\bibinfo{author}{Bradski, G.}
\newblock \bibinfo{journal}{\bibinfo{title}{{The OpenCV Library}}}.
\newblock {\emph{\JournalTitle{Dr. Dobb's Journal of Software Tools}}}
  (\bibinfo{year}{2000}).

\bibitem{Ester1996}
\bibinfo{author}{Ester, M.}, \bibinfo{author}{Kriegel, H.-P.},
  \bibinfo{author}{Sander, J.} \& \bibinfo{author}{Xu, X.}
\newblock \bibinfo{journal}{\bibinfo{title}{{Density-Based Clustering
  Methods}}}.
\newblock {\emph{\JournalTitle{Comprehensive Chemometrics}}}
  \textbf{\bibinfo{volume}{2}}, \bibinfo{pages}{635--654},
  \doiprefix\url{10.1016/B978-044452701-1.00067-3} (\bibinfo{year}{1996}).
\newblock \eprint{10.1.1.71.1980}.

\bibitem{Moshtagh2005}
\bibinfo{author}{Moshtagh, N.}
\newblock \bibinfo{journal}{\bibinfo{title}{{Minimum Volume Enclosing
  Ellipsoids}}}.
\newblock {\emph{\JournalTitle{Convex Optimization}}}
  \textbf{\bibinfo{volume}{111}}, \bibinfo{pages}{112--118}
  (\bibinfo{year}{2005}).

\bibitem{Guimaraes2020}
\bibinfo{author}{Guimaraes, P.~V.} \emph{et~al.}
\newblock \bibinfo{journal}{\bibinfo{title}{{A Data Set of Sea Surface Stereo
  Images to Resolve Space-Time Wave Fields}}}.
\newblock {\emph{\JournalTitle{Scientific Data}}} \textbf{\bibinfo{volume}{7}},
  \bibinfo{pages}{1--12},
  \doiprefix\url{10.12770/af599f42-2770-4d6d-8209-13f40e2c292f}
  (\bibinfo{year}{2020}).

\bibitem{Hastie2009}
\bibinfo{author}{Hastie, T.}, \bibinfo{author}{Tibshirani, R.} \&
  \bibinfo{author}{Friedman, J.}
\newblock \emph{\bibinfo{title}{{The Elements of Statistical Learning}}}
  (\bibinfo{publisher}{Springer International Publishing},
  \bibinfo{year}{2009}), \bibinfo{edition}{second edi} edn.

\bibitem{Shorten2019}
\bibinfo{author}{Shorten, C.} \& \bibinfo{author}{Khoshgoftaar, T.~M.}
\newblock \bibinfo{journal}{\bibinfo{title}{{A survey on Image Data
  Augmentation for Deep Learning}}}.
\newblock {\emph{\JournalTitle{Journal of Big Data}}}
  \textbf{\bibinfo{volume}{6}}, \doiprefix\url{10.1186/s40537-019-0197-0}
  (\bibinfo{year}{2019}).

\bibitem{Simonyan2015}
\bibinfo{author}{Simonyan, K.} \& \bibinfo{author}{Zisserman, A.}
\newblock \bibinfo{journal}{\bibinfo{title}{{Very deep convolutional networks
  for large-scale image recognition}}}.
\newblock {\emph{\JournalTitle{3rd International Conference on Learning
  Representations, ICLR 2015 - Conference Track Proceedings}}}
  \bibinfo{pages}{1--14} (\bibinfo{year}{2015}).
\newblock \eprint{arXiv:1409.1556v6}.

\bibitem{He2016a}
\bibinfo{author}{He, K.}, \bibinfo{author}{Zhang, X.}, \bibinfo{author}{Ren,
  S.} \& \bibinfo{author}{Sun, J.}
\newblock \bibinfo{journal}{\bibinfo{title}{{Identity mappings in deep residual
  networks}}}.
\newblock {\emph{\JournalTitle{Lecture Notes in Computer Science (including
  subseries Lecture Notes in Artificial Intelligence and Lecture Notes in
  Bioinformatics)}}} \textbf{\bibinfo{volume}{9908 LNCS}},
  \bibinfo{pages}{630--645}, \doiprefix\url{10.1007/978-3-319-46493-0_38}
  (\bibinfo{year}{2016}).
\newblock \eprint{1603.05027}.

\bibitem{Szegedy2017}
\bibinfo{author}{Szegedy, C.}, \bibinfo{author}{Ioffe, S.},
  \bibinfo{author}{Vanhoucke, V.} \& \bibinfo{author}{Alemi, A.~A.}
\newblock \bibinfo{journal}{\bibinfo{title}{{Inception-v4, inception-ResNet and
  the impact of residual connections on learning}}}.
\newblock {\emph{\JournalTitle{31st AAAI Conference on Artificial Intelligence,
  AAAI 2017}}} \bibinfo{pages}{4278--4284} (\bibinfo{year}{2017}).
\newblock \eprint{1602.07261}.

\bibitem{Howard2017}
\bibinfo{author}{Howard, A.~G.} \emph{et~al.}
\newblock \bibinfo{title}{{MobileNets: Efficient Convolutional Neural Networks
  for Mobile Vision Applications}} (\bibinfo{year}{2017}).
\newblock \eprint{1704.04861}.

\bibitem{Tan2019}
\bibinfo{author}{Tan, M.} \& \bibinfo{author}{Le, Q.~V.}
\newblock \bibinfo{journal}{\bibinfo{title}{{EfficientNet: Rethinking model
  scaling for convolutional neural networks}}}.
\newblock {\emph{\JournalTitle{36th International Conference on Machine
  Learning, ICML 2019}}} \textbf{\bibinfo{volume}{2019-June}},
  \bibinfo{pages}{10691--10700} (\bibinfo{year}{2019}).
\newblock \eprint{1905.11946}.

\bibitem{Ioffe2015}
\bibinfo{author}{Ioffe, S.} \& \bibinfo{author}{Szegedy, C.}
\newblock \bibinfo{journal}{\bibinfo{title}{{Batch normalization: Accelerating
  deep network training by reducing internal covariate shift}}}.
\newblock {\emph{\JournalTitle{32nd International Conference on Machine
  Learning, ICML 2015}}} \textbf{\bibinfo{volume}{1}},
  \bibinfo{pages}{448--456} (\bibinfo{year}{2015}).
\newblock \eprint{1502.03167}.

\bibitem{Srivastava2014}
\bibinfo{author}{Srivastava, N.}, \bibinfo{author}{Hinton, G.},
  \bibinfo{author}{Krizhevsky, A.}, \bibinfo{author}{Sutskever, I.} \&
  \bibinfo{author}{Salakhutdinov, R.}
\newblock \bibinfo{journal}{\bibinfo{title}{{Dropout: A Simple Way to Prevent
  Neural Networks from Overfitting}}}.
\newblock {\emph{\JournalTitle{Journal of Machine Learning Research}}}
  \textbf{\bibinfo{volume}{15}}, \bibinfo{pages}{1929--1958},
  \doiprefix\url{10.1109/ICAEES.2016.7888100} (\bibinfo{year}{2014}).

\bibitem{Kingma2014}
\bibinfo{author}{Kingma, D.~P.} \& \bibinfo{author}{Ba, J.}
\newblock \bibinfo{title}{{Adam: A Method for Stochastic Optimization}}.
\newblock In \emph{\bibinfo{booktitle}{3rd International Conference for
  Learning Representations}}, \bibinfo{pages}{1--15},
  \doiprefix\url{http://doi.acm.org.ezproxy.lib.ucf.edu/10.1145/1830483.1830503}
  (\bibinfo{address}{San Diego, California}, \bibinfo{year}{2014}).
\newblock \eprint{1412.6980}.

\bibitem{Ruder2016}
\bibinfo{author}{Ruder, S.}
\newblock \bibinfo{title}{{An overview of gradient descent optimization
  algorithms}} (\bibinfo{year}{2016}).
\newblock \eprint{1609.04747}.

\bibitem{tensorflow2015}
\bibinfo{author}{Abadi, M.} \emph{et~al.}
\newblock \bibinfo{title}{{TensorFlow}: Large-scale machine learning on
  heterogeneous systems} (\bibinfo{year}{2015}).
\newblock \bibinfo{note}{Software available from tensorflow.org}.

\bibitem{Nagi2011}
\bibinfo{author}{Nagi, J.} \emph{et~al.}
\newblock \bibinfo{title}{{Max-Pooling Convolutional Neural Networks for
  Vision-based Hand Gesture Recognition}}.
\newblock In \emph{\bibinfo{booktitle}{2011 IEEE International Conference on
  Signal and Image Processing Applications, ICSIPA 2011}},
  \bibinfo{pages}{342--347} (\bibinfo{year}{2011}).

\bibitem{Hanson2004}
\bibinfo{author}{Hanson, J.~L.} \& \bibinfo{author}{Jensen, R.}
\newblock \bibinfo{journal}{\bibinfo{title}{{Wave system diagnostics for
  numerical wave models}}}.
\newblock {\emph{\JournalTitle{8 th International Workshop on Wave Hindcasting
  and Forecasting, Oahu, Hawaii, November}}}  (\bibinfo{year}{2004}).

\bibitem{Large1981}
\bibinfo{author}{Large, W.~G.} \& \bibinfo{author}{Pond, S.}
\newblock \bibinfo{journal}{\bibinfo{title}{{Open Ocean Momentum Flux
  Measurements in Moderate to Strong Winds}}}.
\newblock {\emph{\JournalTitle{Journal of Physical Oceanography}}}
  \doiprefix\url{10.1175/1520-0485(1981)0112.0.CO;2} (\bibinfo{year}{1981}).

\bibitem{Bewley2016}
\bibinfo{author}{Bewley, A.}, \bibinfo{author}{Ge, Z.}, \bibinfo{author}{Ott,
  L.}, \bibinfo{author}{Ramos, F.} \& \bibinfo{author}{Upcroft, B.}
\newblock \bibinfo{journal}{\bibinfo{title}{{Simple online and realtime
  tracking}}}.
\newblock {\emph{\JournalTitle{Proceedings - International Conference on Image
  Processing, ICIP}}} \textbf{\bibinfo{volume}{2016-August}},
  \bibinfo{pages}{3464--3468}, \doiprefix\url{10.1109/ICIP.2016.7533003}
  (\bibinfo{year}{2016}).
\newblock \eprint{1602.00763}.

\bibitem{Holman2007}
\bibinfo{author}{Holman, R.~A.} \& \bibinfo{author}{Stanley, J.}
\newblock \bibinfo{journal}{\bibinfo{title}{{The history and technical
  capabilities of Argus}}}.
\newblock {\emph{\JournalTitle{Coastal Engineering}}}
  \textbf{\bibinfo{volume}{54}}, \bibinfo{pages}{477--491},
  \doiprefix\url{10.1016/j.coastaleng.2007.01.003} (\bibinfo{year}{2007}).

\bibitem{Schwendeman2014}
\bibinfo{author}{Schwendeman, M.}, \bibinfo{author}{Thomson, J.} \&
  \bibinfo{author}{Gemmrich, J.~R.}
\newblock \bibinfo{journal}{\bibinfo{title}{{Wave breaking dissipation in a
  Young Wind Sea}}}.
\newblock {\emph{\JournalTitle{Journal of Physical Oceanography}}}
  \textbf{\bibinfo{volume}{44}}, \bibinfo{pages}{104--127},
  \doiprefix\url{10.1175/JPO-D-12-0237.1} (\bibinfo{year}{2014}).

\bibitem{Guimaraes2018}
\bibinfo{author}{Guimaraes, P.~V.}
\newblock \emph{\bibinfo{title}{{Sea surface and energy dissipation}}}.
\newblock Ph.D. thesis, \bibinfo{school}{Universit{\`{e}} de Bretagne Loire}
  (\bibinfo{year}{2018}).

\bibitem{Duncan1981}
\bibinfo{author}{Duncan, J.~H.}
\newblock \bibinfo{journal}{\bibinfo{title}{{An Experimental Investigation of
  Breaking Waves Produced by a Towed Hydrofoil}}}.
\newblock {\emph{\JournalTitle{Proceedings of the Royal Society A:
  Mathematical, Physical and Engineering Sciences}}}
  \textbf{\bibinfo{volume}{377}}, \bibinfo{pages}{331--348},
  \doiprefix\url{10.1098/rspa.1981.0127} (\bibinfo{year}{1981}).

\bibitem{2020SciPy-NMeth}
\bibinfo{author}{{Virtanen}, P.} \emph{et~al.}
\newblock \bibinfo{journal}{\bibinfo{title}{{SciPy 1.0: Fundamental Algorithms
  for Scientific Computing in Python}}}.
\newblock {\emph{\JournalTitle{Nature Methods}}}
  \doiprefix\url{https://doi.org/10.1038/s41592-019-0686-2}
  (\bibinfo{year}{2020}).

\bibitem{Russakovsky2015}
\bibinfo{author}{Russakovsky, O.} \emph{et~al.}
\newblock \bibinfo{journal}{\bibinfo{title}{{ImageNet Large Scale Visual
  Recognition Challenge}}}.
\newblock {\emph{\JournalTitle{International Journal of Computer Vision}}}
  \textbf{\bibinfo{volume}{115}}, \bibinfo{pages}{211--252},
  \doiprefix\url{10.1007/s11263-015-0816-y} (\bibinfo{year}{2015}).
\newblock \eprint{1409.0575}.

\bibitem{Selvaraju2020}
\bibinfo{author}{Selvaraju, R.~R.} \emph{et~al.}
\newblock \bibinfo{journal}{\bibinfo{title}{{Grad-CAM: Visual Explanations from
  Deep Networks via Gradient-Based Localization}}}.
\newblock {\emph{\JournalTitle{International Journal of Computer Vision}}}
  \textbf{\bibinfo{volume}{128}}, \bibinfo{pages}{336--359},
  \doiprefix\url{10.1007/s11263-019-01228-7} (\bibinfo{year}{2020}).
\newblock \eprint{1610.02391}.

\bibitem{RFB15a}
\bibinfo{author}{Ronneberger, O.}, \bibinfo{author}{P.Fischer} \&
  \bibinfo{author}{Brox, T.}
\newblock \bibinfo{title}{U-net: Convolutional networks for biomedical image
  segmentation}.
\newblock In \emph{\bibinfo{booktitle}{Medical Image Computing and
  Computer-Assisted Intervention (MICCAI)}}, vol. \bibinfo{volume}{9351} of
  \emph{\bibinfo{series}{LNCS}}, \bibinfo{pages}{234--241}
  (\bibinfo{publisher}{Springer}, \bibinfo{year}{2015}).
\newblock \bibinfo{note}{(available on arXiv:1505.04597 [cs.CV])}.

\bibitem{Donelan2010}
\bibinfo{author}{Donelan, M.~A.}, \bibinfo{author}{Haus, B.~K.},
  \bibinfo{author}{Plant, W.~J.} \& \bibinfo{author}{Troianowski, O.}
\newblock \bibinfo{journal}{\bibinfo{title}{{Modulation of short wind waves by
  long waves}}}.
\newblock {\emph{\JournalTitle{Journal of Geophysical Research: Oceans}}}
  \textbf{\bibinfo{volume}{115}}, \bibinfo{pages}{1--12},
  \doiprefix\url{10.1029/2009JC005794} (\bibinfo{year}{2010}).

\end{thebibliography}

\section*{Acknowledgements}

This work benefited from France Energies Marines and State financing managed by the National Research Agency under the Investments for the Future program bearing the reference numbers ANR-10-IED-0006-14 and ANR-10-IEED-0006-26 for the projects DiME and CARAVELE.

\section*{Author contributions statement}

C.E.S. idealized and implemented the method, C.E.S. drafted the manuscript, P.V.G., F.L., J.F.F. and R. D. conducted the field experiments, C.E.S. and P.V.G. analyzed the data, J.F. provided funding. All authors reviewed the manuscript. 

\section*{Additional information}

Data, pre-trained networks and auxiliary programs necessary to reproduce the results in this paper are available at: \url{https://github.com/caiostringari/deepwaves}.

\section*{Competing Interests Statement}

The authors declare no known competing interests.

\end{document}